\documentclass[aps,a4paper,showpacs,prc,floatfix,twocolumn,amsmath,amssymb,nofootinbib]{revtex4-1}
\usepackage{graphicx}
\usepackage{epsfig}
\usepackage{ulem}
\usepackage{morefloats}
\usepackage{rotating}
\usepackage{color,ulem}
\usepackage{graphicx}

\begin{document}

\title{A Nonequilibrium Information Entropy Approach to Ternary Fission of Actinides}

\author{G. R\"{o}pke$^1$}
\email{gerd.roepke@uni-rostock.de}
\author{J. B. Natowitz$^2$}
\email{natowitz@comp.tamu.edu}
\author{H. Pais$^3$}
\email{hpais@uc.pt}
\affiliation{$^1$Institut f\"{u}r Physik, Universit\"{a}t Rostock, D-18051 Rostock, Germany. \\
$^2$Cyclotron Institute, Texas A\&M University, College Station, Texas 77843, USA. \\
$^3$CFisUC,  Department of Physics, University of Coimbra, 3004-516 Coimbra, Portugal.}

\date{\today}

\begin{abstract}
Ternary fission of actinides probes the state of the nucleus at scission. 
Light clusters are produced in space and time very close to the scission point.
Within the nonequilibrium statistical operator method, a generalized Gibbs distribution is constructed 
from the information given by the observed yields of isotopes. Using this relevant statistical operator, 
yields are calculated taking excited states and continuum correlations into account, 
in accordance with the virial expansion of the equation of state.
Clusters with mass number $A \le 10$ are well described
 using the nonequilibrium generalizations of temperature and chemical potentials. 
Improving the virial expansion, in-medium effects may become of importance in determining the contribution of weakly bound states 
and continuum correlations to the intrinsic partition function. 
Yields of larger clusters, which fail to reach this quasi-equilibrium form of the relevant distribution, 
are described by nucleation kinetics, and a saddle-to-scission relaxation time of about 7000 fm/c is inferred. 
Light charged particle emission, described by reaction kinetics and virial expansions, may therefore be regarded 
as a very important tool to probe the nonequilibrium time evolution of actinide nuclei during fission.
\end{abstract}

\pacs{21.65.-f, 21.60.Jz, 25.70.Pq, 26.60.Kp}

\maketitle

Nuclear fission, discovered eighty years ago, remains an exciting field of research. 
In the last few decades, an amazing progress has been realized with respect to experimental investigations and phenomenology, as well as in theoretical treatments such as time-dependent 
Hartree-Fock-Bogoliubov (TDHFB) or time-dependent superfluid local density approximation, generator coordinate methods, 
and other techniques \cite{Bulgac19a,Bertsch19}, but basic concepts are still open for discussion.
A full microscopic description is still lacking, and it remains a challenge to present state quantum many-body theory.
 For a recent review on studies of thermal neutron induced ($n_{\rm th}$,f), and spontaneous fission (sf) of actinides, as well as a discussion  on open theoretical questions, the reader should refer to \cite{Bulgac20,Bender20}.

From a theoretical point of view, the fission process can be generally described via a picture in which the deforming nucleus, 
following a dynamic path, subject to fluctuations, crosses a saddle point where the nascent fission fragments are formed. 
The deformed dumbbell-like system, consisting of two main fragments and the connecting neck region,
then evolves toward the scission point where the rupture occurs.
The saddle-to-scission time is  estimated as $\tau_{s \to s}\approx O(10^3 - 10^4)$ fm/c  \cite{Bulgac19a}.
Characterization of the system's properties and its time evolution,
and the description of dissipation during this process  remains nowadays a significant problem.
However, dissipative dynamics has been applied to describe the non-adiabatic 
evolution from the saddle point to scission, see \cite{Bulgac20},  though a 
rigorous treatment of the scission process is still unavailable.

\begin{table*}
\begin{center}
\hspace{0.5cm}
 \begin{tabular}{|c|c|c|c|c|c|c|c|} 
\hline
isotope  &$R^{\rm vir}_{A,Z}(1.3)$& $^{233}$U($n_{\rm th}$,f) & $^{235}$U($n_{\rm th}$,f) & $^{239}$Pu($n_{\rm th}$,f) & $^{241}$Pu($n_{\rm th}$,f) & $^{248}$Cm(sf) &
 $^{252}$Cf(sf) \\
\hline
 $\lambda_T$ [MeV]	&- & 1.24177 	&  1.21899 	& 1.3097	& 1.1900		& 1.23234	& 1.25052	\\
$\lambda_n$ 	[MeV]	&-& -3.52615 	&  -3.2672	& -3.46688	& -3.02055	& -2.92719	& -3.1107	\\
$\lambda_p$ [MeV]	&-& -15.8182 	&  -16.458	& -16.2212	& -16.6619 	& -16.7798	& -16.7538	\\
\hline
$^1$n		 	&-& 560012  	&  1.409e6	& 722940	& 1.8579e6	& 1.606e6	& 1.647e6	\\
$^1$H			&-& 28.131 	&  28.16	& 42.638 	& 19.52		& 21.079 	& 30.096	\\
$^2$H$^{\rm obs}$	&-& 41		&  50		& 69		& 42 		& 50		& 63		\\
$^2$H 			&0.973& 40.986 	&  49.76	& 68.632	& 41.563	& 49.533 	& 61.579 	\\
$^3$H$^{\rm obs}$	&-& 460		&  720		& 720 		& 786		& 922 		& 950		\\
$^3$H			&0.998& 457.27 	&  715.29	& 714.79	& 780.39	& 913.76 	& 943.12	\\
$^4$H			&0.0876& 2.7772  	&  4.97		& 5.627		& 6.057 	& 8.742 	& 8.219 	\\
$^3$He			&0.997& 0.0124	&  0.0076	& 0.0235 	& 0.00431 	& 0.00645 	& 0.00933	\\
$^4$He$^{\rm obs}$	&-& 10000  	&  10000	& 10000		& 10000		& 10000		& 10000		\\
$^4$He			&1& 8858.46 	&  8706.1	& 8615.7	& 8556.9	& 8313.98	& 8454.0	\\
$^5$He			&0.689& 1130.75 	&  1289.04	& 1374.7	& 1439.0	& 1680.75	& 1540.9	\\
$^6$He$^{\rm obs}$	&-& 137		&  191		& 192 		& 260 		& 354 		& 270 		\\
$^6$He			&0.933& 115.89  	&  158.98	& 159.01 	& 211.68	& 276.96	& 222.4		\\
$^7$He			&0.876& 21.262 	&  33.997	& 35.983 	& 51.742 	& 80.634 	& 58.16		\\
$Y_{^6{\rm He}}^{\rm obs}/Y_{^6{\rm He}}^{\rm final,vir}$ &-& 0.9989	&  0.9897	& 0.9846	& 0.9869 	& 0.9899	& 0.9622	\\
$^8$He$^{\rm obs}$	&-& 3.6		&  8.2		& 8.8		& 15		& 24		& 25		\\
$^8$He			&0.971& 3.4725 	&  6.764	& 6.4095 	& 12.481 	& 21.280 	& 13.32		\\
$^9$He			&0.255& 0.047077 	&  0.105	& 0.111		& 0.219 	& 0.455 	& 0.258 	\\
$Y_{^8{\rm He}}^{\rm obs}/Y_{^8{\rm He}}^{\rm final,vir}$ &-& 1.0229 	&  1.1936	& 1.3496	& 1.1811	& 1.1042 	& 1.8409	\\
$^8$Be			&1.07& 5.7727 	&  2.594 	& 5.147	& 2.188 	& 2.819 	& 2.544		\\
\hline
 \end{tabular}
\caption{
Lagrange parameters $\lambda_i$, observed yields $Y^{\rm obs}_{A,Z}$ \cite{Valskii2,Koester,Kopatch02} and primary yields $Y^{\rm rel, vir}_{A,Z}$ (virial approximation)  from ternary fission   $^{233}$U($n_{\rm th}$,f), $^{235}$U($n_{\rm th}$,f), $^{239}$Pu($n_{\rm th}$,f), $^{241}$Pu($n_{\rm th}$,f), $^{248}$Cm(sf), and $^{252}$Cf(sf) (lines denoted by the superscript 'obs') as well as relevant (primary) yields $Y^{\rm rel, vir}_{A,Z}$ for H, He nuclei. The prefactor $R^{\rm vir}_{A,Z}(\lambda_T)$ at $\lambda_T = 1.3$ MeV which represents the intrinsic partition function is also given. The final yields $Y_{^6{\rm He}}^{\rm final,vir}=Y^{\rm rel, vir}_{^6{\rm He}}+Y^{\rm rel, vir}_{^7{\rm He}}$ 
and $Y_{^8{\rm He}}^{\rm final,vir}=Y^{\rm rel, vir}_{^8{\rm He}}+Y^{\rm rel, vir}_{^9{\rm He}}$ are compared to the observed yields. For more details see the Supplementary material \cite{Suppl}.}
\label{Tab:HHevir}
\end{center}
\end{table*}

 A few signals can be used to obtain information about the scissioning nucleus, such as  
the mass and energy distributions of the two fission fragments, 
and the multiplicities of the emitted particles, which are primarily  neutrons,  and of  $\gamma$-radiation, 
that can be utilized to characterize excitation energies and spins.
Although the use of concepts, like the temperature, which is defined for thermodynamic equilibrium,
might not be well-founded, approaches introducing concepts from statistical physics, 
to characterize the distribution of emitted particles, are often employed.
For instance, in Refs.~\cite{Granier,Iwamoto,Kocak}, the prompt fission neutron spectra of different actinides are analyzed with temperature-like parameters of the order of 1 MeV,
and in Refs.~\cite{Choudhury19,Oberstedt13,Gatera17,Lebois15,Talou15,Makii19}, the analysis of prompt fission $\gamma$-ray spectra for actinides also suggests a temperature-like parameter of the same order.
However, one has to separate prompt emission  from later emission,
that occurs during the de-excitation of the fission fragments, and this presents a  problem, since it is not easy to identify observables which may be directly associated
with the neck region at scission.

One such signature is the emission of light clusters observed in ternary fission processes, 
see, e.g.,
\cite{Koester,KoesterPhD,Kopatch02} and references given there. 
A light cluster, most often $^4$He, is emitted in a direction perpendicular to
the symmetry axis defined by the two fission products, which have mass numbers distributed near half that of the  fissioning nucleus.
Ternary fission yields of a series of light isotopes $\{A,Z\}$ and energies have been measured for a number of different actinide nuclei,  in particular
$^{233}$U($n_{\rm th}$,f), $^{235}$U($n_{\rm th}$,f), $^{239}$Pu($n_{\rm th}$,f), $^{241}$Pu($n_{\rm th}$,f), 
$^{248}$Cm(sf), and $^{252}$Cf(sf), see Refs. \cite{Valskii2,KoesterPhD}. 
Data for these observed yields $Y^{\rm obs}_{A,Z}$, normalized to $^4$He$^{\rm obs}= 10000$, are presented in Tab. \ref{Tab:HHevir}. 
The investigation of ternary fission has the advantage 
that it is directly related to the scission process, and that it can be localized  in the neck region.

As known from $\alpha$-decay studies, a mean-field approach like TDHFB has problems describing the formation of clusters. 
For ternary fission, parametrizations of the measured yields employing
a statistical distribution with a temperature-like parameter $T \approx 1$ MeV, 
see \cite{Valskii2,andronenko,Lestone05}, have been explored.  However, 
any interpretation of the detected yields by a simple nuclear statistical equilibrium (NSE) model, see Eq. (\ref{Y0}) below, 
faces some problems. The observed yields seen in the detector contain contributions from  
decaying excited states and resonances so that the observed yield distribution differs from the primary distribution at the time of scission. 
In addition, yields of the larger light isotopes, $A \ge 10$, 
are clearly overestimated by the simple NSE statistical equilibrium distribution \cite{Valskii2}. 
Modifications have been proposed \cite{Sara} based on nucleation theory. 
In  Ref. \cite{rnp20}, a nonequilibrium approach was used to discuss the observed yields of isotopes with $Z \le 2$ for the spontaneous fission of $^{252}$Cf. 
Chemical equilibrium constants were recently derived \cite{natowitz20}  for the fission reaction $^{241}$Pu($n_{\rm th}$,f) accounting for in-medium effects. Using those constants,
 the predicted yields for increasing $A$ are larger than the measured ones, as already observed in \cite{Sara}. 
 
In this work, we use ternary fission data to investigate the scission process. In particular, we extend the nonequilibrium approach, already presented in Ref.~\cite{rnp20} but with $Z \le 2$, considering partial intrinsic partition functions including continuum contributions 
on the level of quantum virial expansions. We present results for all the actinides given in Table \ref{Tab:HHevir}  for which the necessary set of yields is available, and we derive critical values of parameters that are relevant for nucleation kinetics. 

We describe fission as a nonequilibrium process, using the method of the nonequilibrium statistical operator 
(NSO), 
\begin{equation}
\label{rhoZ}
\rho(t)=\lim_{\epsilon \to 0} \epsilon \int_{-\infty}^{ t} d t' e^{-\epsilon ( t- t')}
e^{-\frac{i}{\hbar} H ( t- t')}\rho_{\rm rel}( t')e^{\frac{i}{\hbar} H ( t- t')}.
\end{equation}
 The NSO $\rho(t)$ \cite{ZMR} is a solution of the von Neumann equation 
with boundary conditions
characterizing the state of the system in the past, as expressed by the relevant statistical operator.
This relevant statistical operator $\rho_{\rm rel}(t)$ is constructed from known averages using information theory.
As it is well known from statistical physics, 
the relevant distribution is determined from the maximum of information entropy under given constraints, which are represented by Lagrange multipliers $\lambda_i$. 
A minimum set of relevant observables consists of the conserved observables, energy $H$, and the numbers $N_\tau$  of neutrons and protons ($\tau = n, p$).
The solution is the generalized Gibbs distribution 
\begin{equation}
\rho_{\rm rel} \propto \exp[-(H-\lambda_n N_n-\lambda_p N_p)/\lambda_T] \, .
\end{equation}
Note that these Lagrange multipliers $\lambda_i$, which are in general time-dependent, 
are not identical with the equilibrium parameters $T$ and  $\mu_\tau$, but may be considered as nonequilibrium generalizations of the temperature and chemical potentials. 
Only when the system is in thermodynamic equilibrium, can the information entropy  be unambigously identified with the thermodynamic entropy, 
and from this we define the
quantities  $T$ and   $\mu_\tau$  with the known properties. 
Note that the NSO allows the possibility of including other relevant observables, such as the pair amplitude in the superfluid state, or
the occupation numbers of the quasiparticle states to derive kinetic equations and calculating reaction rates \cite{ZMR}.

Typically, in a variational problem, we have to replace the Lagrange multipliers 
$\lambda_i$ by given mean values \cite{rnp20}. 
In contrast to the noninteracting system (ideal quantum gases), where the equilibrium solutions are the well-known equations of state, 
the interaction in the Hamiltonian leads to a many-particle problem which can be treated 
with the methods of quantum statistics. 
(Note that the mathematical concepts developed in equilibrium quantum statistics can also used 
for the generalized Gibbs state $\rho_{\rm rel}$.)
We perform a cluster expansion for the interacting system \cite{R20}, and partial densities of different clusters $\{A,Z\}$ are introduced. 
The relevant yields $Y^{\rm rel, vir}_{A,Z}$ in the virial approximation are calculated as 
\begin{eqnarray} 
\label{Y0}
Y^{\rm rel, vir}_{A,Z} &\propto & R^{\rm vir}_{A,Z}(\lambda_T)\, g_{A,Z} \left(\frac{2 \pi \hbar^2}{A m \lambda_T}\right)^{-3/2} \times  \nonumber \\
&&e^{(B_{A,Z}+(A-Z) \lambda_n+Z  \lambda_p)/ \lambda_T}
\end{eqnarray}  
(nondegenerate limit), where $B_{A,Z}$ denotes the (ground state) binding energy and $g_{A,Z}$ the degeneracy \cite{nuclei}.
The prefactor 
\begin{eqnarray}
R^{\rm vir}_{A,Z}(\lambda_T)=1+\sum^{\rm exc}_i [g_{A,Z,i}/g_{A,Z}] e^{-E_{A,Z,i}/\lambda_T}
\end{eqnarray}
is related to the intrinsic partition function of the cluster $\{A,Z\}$.
The summation is performed over all excited states of excitation energy $E_{A,Z,i}$ and degeneracy $g_{A,Z,i}$  \cite{nuclei}, 
which decay to the ground state.
Also, the continuum contributions are included in the virial expression. For instance, 
the Beth-Uhlenbeck formula expresses the contribution of the continuum to the intrinsic partition function via the scattering phase shifts, see \cite{SRS,HS}. 
For $R^{\rm vir}_{A,Z}(\lambda_T)=1$, the simple NSE is obtained, 
i.e., neglecting the contribution of all excited states inclusive continuum correlations.

In the low-density limit, virial expansions of the intrinsic partition functions of the channel $\{A,Z\}$ have been obtained \cite{SRS,HS,R20} for $^2$H, $^4$H, $^5$He, $^8$Be using the measured phase shifts in the corresponding channels.
The values are given in  the second column of Tab. \ref{Tab:HHevir} for $\lambda_T=1.3$ MeV.  Well-bound states with energy of the continuum edge large compared to $\lambda_T$ have only 
a weak contribution of the continuum states so that $R^{\rm vir}_{A,Z}(\lambda_T)\approx 1$, if no further excited states are present. 
An interpolation formula, which relates the prefactor $R^{\rm vir}_{A,Z}(\lambda_T)$ to the energy 
of the edge of continuum, is given in \cite{R20}, and the corresponding estimates of the prefactor for the He isotopes  with $6 \le A \le 9$ are also shown in Tab. \ref{Tab:HHevir}.

Within the NSO approach (\ref{rhoZ}), the relevant distribution $\rho_{\rm rel}$ serves as 
initial condition to solve the von Neumann equation for $\rho(t)$
describing the evolution of the system according to the system Hamiltonian.
The relevant (primary) distribution $Y^{\rm rel}_{A,Z}(\lambda_T)$ contains stable and unstable states of nuclei, 
as well as correlations in the continuum (e.g. resonances). 

The concept of introducing the relevant primary yield distribution according to the NSO is supported by several experimental observations, among these are
the observation of $^5$He and $^7$He emission \cite{Kopatch02}. For $Z >2$, as it is discussed below,  3.368 MeV $\gamma$ rays 
from the first excited state of $^{10}$Be have been observed \cite{Daniel04}, and  excited states of $^8$Li at 2.26 MeV excitation energy have been reported
in Ref.~\cite{Kopatch02}.
Also of interest are the inferred data for $^8$Be and 
$^7$Li$_{7/2^-}$ observed in \cite{Jesinger05}, which cannot be described with the NSE 
but demand a treatment with continuum states.

The relevant distribution evolves dynamically to the final, observed yields according to the von Neumann equation. 
This process is described by reaction kinetics, and the NSO allows calculation of the reaction rates \cite{ZMR}. 
Here, we approximate this process by the feeding of  the observed states 
from the primary states occurring in the relevant distribution.
For example, for $Z \le 2$, the final yields are related to the primary, relevant yields as 
$Y_{^3{\rm H}}^{\rm final}=Y^{\rm rel}_{^3{\rm H}}+Y^{\rm rel}_{^4{\rm H}}$, $\,\,
 Y_{^4{\rm He}}^{\rm final}=Y^{\rm rel}_{^4{\rm He}}+Y^{\rm rel}_{^5{\rm He}}+2 Y^{\rm rel}_{^8{\rm Be}}$, $\,\, 
Y_{^6{\rm He}}^{\rm final}=Y^{\rm rel}_{^6{\rm He}}+Y^{\rm rel}_{^7{\rm He}}$, 
and $ Y_{^8{\rm He}}^{\rm final}=Y^{\rm rel}_{^8{\rm He}}+Y^{\rm rel}_{^9{\rm He}}$.

In this work, to construct the relevant distribution $\rho_{\rm rel}$ from an information theoretical approach,  
we use the least squares method, see \cite{Sara}, to reproduce the observed yields. 
We calculate the primary distribution $Y^{\rm rel,vir}_{A,Z}$ using the 
intrinsic partition function in the virial form, i.e. using the excited states and scattering phase shifts 
neglecting in-medium corrections. The optimum values of the Lagrange parameters $\lambda_i$
are given in Tab. \ref{Tab:HHevir} for the different ternary fissioning actinides.
Of interest is the dependence of $\lambda_i$ from $\{A, Z\}$ of the parent actinide nucleus 
 \cite{Lestone05,andronenko}.
The current accuracy of the experimental data is not sufficient to determine significant trends.

The measured total yields of H and He isotopes are nearly perfectly reproduced by the corresponding sums of primary yields.
The yield of $^6$He is slightly overestimated
by $Y_{^6{\rm He}}^{\rm final,vir}$. 
In contrast, the yield of $^8$He is underestimated
by $Y_{^8{\rm He}}^{\rm final,vir}$. 
Both ratios $Y_{^6{\rm He}}^{\rm obs}/Y_{^6{\rm He}}^{\rm final,vir}$ and $Y_{^8{\rm He}}^{\rm obs}/Y_{^8{\rm He}}^{\rm final,vir}$ 
are presented in Tab. \ref{Tab:HHevir}.
Presently, the relevant distribution  $Y^{\rm rel, vir}_{A,Z}$ does not take in-medium effects, in particular Pauli blocking, into account. 
Medium modifications are more effective for weakly bound clusters. 
As proposed in \cite{rnp20} for $^{252}$Cf(sf), a stronger reduction of the yield of $^6$He$^{\rm obs}$ 
compared to $^8$He$^{\rm obs}$ may be related to the very low binding energy (0.975 MeV)
of the $^6$He nucleus below the $\alpha + 2 n$ threshold. The suppression of $^6$He$^{\rm obs}$ appears for all considered systems and may be considered as a signature of the Pauli blocking. Pauli blocking is determined by the medium 
surrounding the cluster, and an estimate of the corresponding neutron density was given in \cite{rnp20}.
However, to address the problem, precise experimental data are needed.
Experimental studies are still scarce, and the
data are often not consistent \cite{Mutterer08,Vermote10}. 
Unbound nuclei such as $^5$He should be very sensitive to medium modifications. 
The virial expression for the intrinsic partition function is known \cite{HS}, 
and the corresponding primary yields are given in Tab. \ref{Tab:HHevir}. 
Fortunately, in the case of  $^{252}$Cf, the primary yields of $^5$He and $^7$He have been measured \cite{Kopatch02},
and the  value $Y^{\rm obs}_{^5{\rm He}}=1736(274)$ has been obtained. 
In principle, because of the medium modifications, the different cluster states may serve as a probe 
to determine the neutron density in the neck region at scission, 
but the uncertainties are still rather large. 

Nuclei with $Z >2$ are also observed in ternary fission. 
A detailed measurement of the yields of isotopes up to $^{30}$Mg has been made  for 
$^{241}$Pu($n_{\rm th}$,f)  \cite{Koester,KoesterPhD}. Extended sets of data for $Z >2$ are also measured for $^{235}$U($n_{\rm th}$,f)
and $^{245}$Cm($n_{\rm th}$,f) \cite{KoesterPhD}. We extend our analysis of the measured data up to $Z=6$ 
using the relevant distribution, see \cite{Suppl}. 
In general, the neutron separation energy $S_n$, for each isotope, is adopted as the threshold energy for the continuum, but cluster decay is also possible, e.g.,  
$^6$Li $\to \alpha + d$, $^7$Li $\to \alpha + t$, $^7$Be $\to \alpha + h$, $^8$Be $\to 2 \alpha $, $^{10}$B $\to \alpha + ^6$Li, etc. 
In some cases, such as $^6$He, $^8$He, $^{11}$Li, two-neutron separation determines the threshold.
To estimate the continuum correlation, the interpolation 
$R^{\rm vir}_{A,Z}(\lambda_T)$ \cite{rnp20} was used at the corresponding binding energy $E^{\rm thresh}_{AZ}-E_{A,Z,i}$ of the (ground state or excited) cluster. 
The final yields $Y^{\rm final, vir}_{A,Z}$ are calculated as sums of the feeding contributions, see \cite{Suppl}.

The question arises whether global Lagrange parameters $\lambda_T,\lambda_n,\lambda_p$ exist, which are valid for all $Z$, 
as expected for matter in thermodynamic equilibrium. Before we discuss this question, 
we present a calculation with the relevant distribution given above, employing only three Lagrange parameters $\lambda_i$, 
but taking also Li isotopes into account.
A least squares fit of final yields $Y^{\rm final, vir}_{A,Z}$ to $Y^{\rm obs}_{A,Z}$ 
for $^2$H, $^3$H, $^4$He, $^8$He, $^7$Li,  $^8$Li,  $^9$Li has been performed. 
The accuracy of the fit increases since $^6$He and $^{11}$Li are not included. 
Both are weakly bound systems for which medium effects and dissolution may become of relevance, as discussed above. 
Again we emphasize that in-medium corrections are not included in the present calculation. 
The Lagrange parameter values
$\hat \lambda_T=1.2023$ MeV, $\hat \lambda_n= -2.9981$ MeV, $\hat \lambda_p=-16.6285$ MeV are obtained. 
There are only minimal changes compared to those derived from the fit of Tab. \ref{Tab:HHevir} 
for $^{241}$Pu($n_{\rm th}$,f), and we conclude that our approach can also reproduce the yields of  
isotopes with $Z > 2$.

\begin{figure}[h]
\centerline{
\,\,\,\,\,\includegraphics[width=280pt,angle=0]{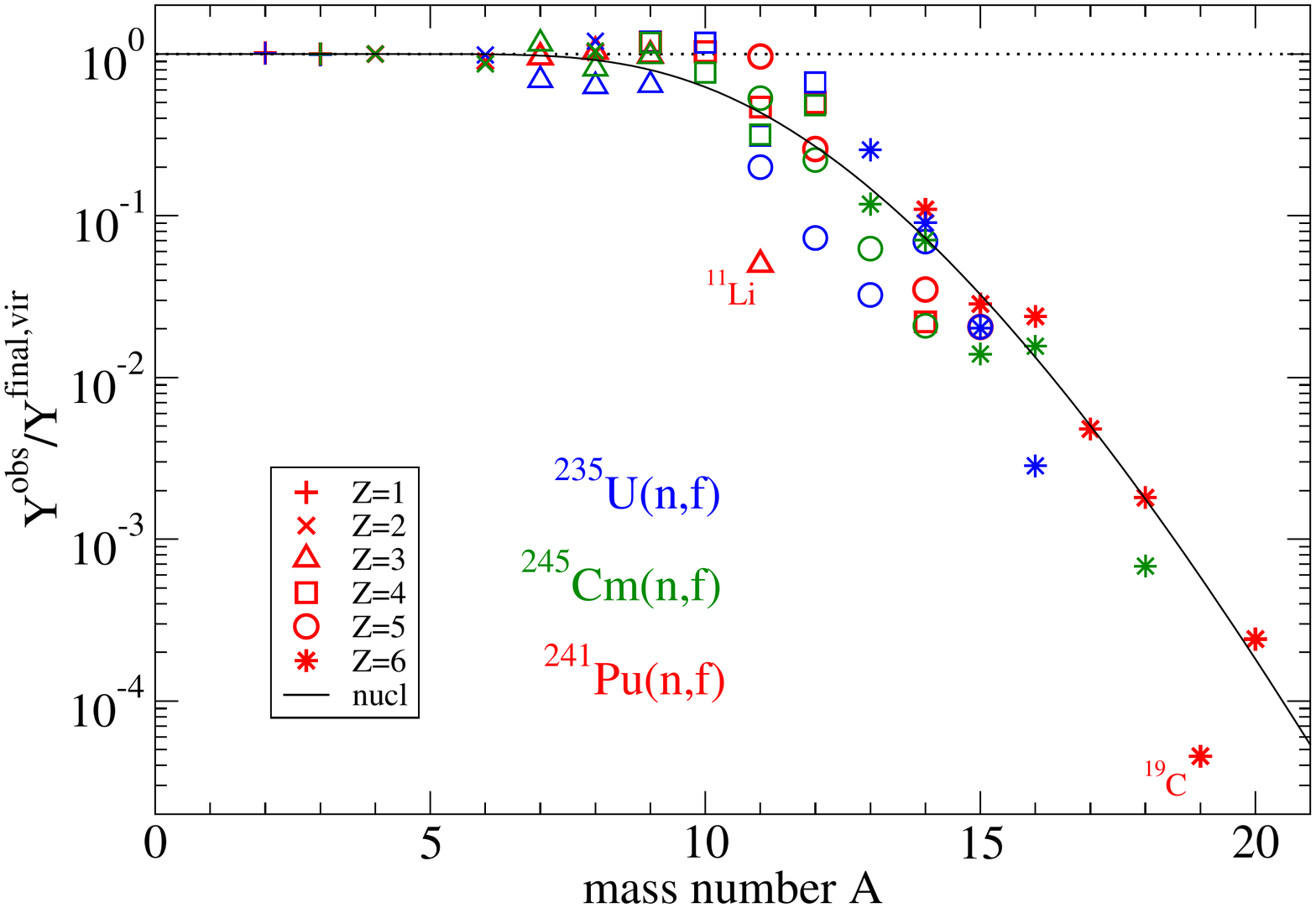}}
\caption{
Ternary fission of $^{235}$U($n_{\rm th}$,f) (blue), $^{245}$Cm($n_{\rm th}$,f) (green), and $^{241}$Pu($n_{\rm th}$,f) (red): Ratio $Y^{\rm obs}_{A,Z}$/Y$^{\rm final, vir}_{A,Z}$ as function of the mass number $A$. Isotopes with $Z \le 6$ are shown. Black full line: Fit of nucleation kinetics 
(\ref{nucleation}) to the data of $^{241}$Pu($n_{\rm th}$,f). Data tables are given in \cite{Suppl}.}
\label{fig:letter1}
\end{figure}

Using these Lagrange parameter values $\hat \lambda_i$ and considering all observed data for isotopes with $Z\le 6$, 
the ratio $Y^{\rm obs}_{A,Z}$/$Y^{\rm final, vir}_{A,Z}$ is shown as a function of the mass number $A$ in Fig. \ref{fig:letter1}.
Surprisingly, yields of $^9$Be, $^{10}$Be, $^{11}$B are also well reproduced.  For $A \ge 11$, the calculations overestimate the observed yields, and the ratios decrease strongly, 
starting around $A=10$.

An explanation of the decrease has been given in \cite{Sara} using nucleation theory.
Whereas small clusters are already in the quasi-equilibrium distribution $Y^{\rm rel}_{A,Z}$, larger clusters need more formation time so that the observed yields are smaller than those predicted by the relevant distribution.
From reaction kinetics, the expression 
\begin{equation}
Y^{\rm obs}_{A,Z}/Y^{\rm final, vir}_{A,Z}=\frac{1}{2} {\rm erfc}\left[b(\tau) (A^{1/3}-a(A_c,\tau))\right]
\label{nucleation}
\end{equation}
is obtained, see \cite{Sara},
where $b(\tau)= (27.59 \,{\rm MeV}/ \lambda_T)^{1/2} \times (1-e^{-2 \tau})^{-1/2}$ and $a(A_c,\tau)=A_c^{1/3}(1-e^{-\tau})+e^{-\tau}$.
With 
$ \lambda_T=1.2$  MeV, the least squares fit to the data of $^{241}$Pu($n_{\rm th}$,f) (black line in Fig. \ref{fig:letter1}) gives
$\tau = 1.5406,\,\,A_c=16.143$. Then, 
$c\tau_{s \to s}=\tau A_c^{2/3} \lambda_T^{1/2}/(3.967\, \rho)$, and with $\rho=4 \times 10^{-4}$ fm$^{-3}$ \cite{Sara} follows $\tau_{s \to s}=6793$ fm/$c$.
This time scale supports the slow evolution from saddle to scission proposed recently as a dissipative process \cite{Bulgac19a,Bulgac20}.

The strong reduction of isotopes  $A > 10$ compared to estimates of a statistical model is also seen in \cite{Valskii2}.
In addition, the overestimate of $^5$He is shown. The correct treatment of continuum states proposed in this letter removes this discrepancy. 
In addition, the yields of weakly bound clusters $^{11}$Li, $^{19}$C  are strongly overestimated, see \cite{Koestera}. 
A reason may be the shift of the binding energy due to in-medium effects. If the density is larger than the Mott density, 
the bound states are dissolved. Bond states with threshold energies 
below  or near 1 MeV include also $^6$He, $^{11}$Be,  $^{14}$Be,  $^{14}$B, $^{15}$C. The yields of all these isotopes are overestimated. 
This may be considered as indication of in-medium effects (Pauli blocking) leading to a shift and possibly the dissolution of the cluster. 
This possibility should be considered when more accurate data are available.

Yields of ternary fission of  $^{241}$Pu($n_{\rm th}$,f)  were also calculated within the 
evaporation-based surface-plus-window dissipation model  \cite{Lestone05,Lestone08}, 
where the yields of light clusters are obtained considering the height of the Coulomb barrier as a dynamical variable. 
The Coulomb interaction has been treated as mean field. 
To improve  the liquid drop model used in that phenomenological approach, an adequate theory of ternary fission 
should be based on the TDHFB solutions. It should be pointed out that the neutrons and protons have extended wave functions and are not correctly described by the distribution (\ref{Y0}), see \cite{rnp20}. The Coulomb interaction contributes to the Hamiltonian also in our approach which will be extended to the treatment of inhomogeneous systems. As pointed out in Ref.  \cite{Bulgac19a}, the dynamics near scission is strongly dissipative and existing versions of TDDFT are not adequate since they are lacking fluctuations in collective coordinates. 

In conclusion, ternary fission is a dissipative process, 
and understanding fission of actinide nuclei is of fundamental interest to work out nonequilibrium statistical physics.
Emitted light isotopes are described by a nonequilibrium distribution, excited states and continuum correlations are taken into account on the level of virial expansions.
In-medium effects and nucleation kinetics are interesting aspects to reconstruct the nonequilibrium distribution.
An improved description of the influence of mean-field effects may be obtained from TDHFB calculations and similar approaches, but cluster formation remains problematic in any mean-field theory.
Ternary fission is an  outstanding signal to explore the fission process, and more consistent and accurate data are necessary to work out a complete description of  the ternary fission process within non-equilibrium quantum statistics.
\\

{\bf Acknowledgments}\\
 
This work was supported by the United States Department of Energy under Grant \# DE-FG03- 93ER40773, by the German Research Foundation (DFG), Grant \# RO905/38-1, the FCT (Portugal) Projects No. UID/FIS/04564/2019 and UID/FIS/04564/2020, and POCI-01-0145-FEDER-029912, and by PHAROS COST Action CA16214. H. P. acknowledges the grant CEECIND/03092/2017 (FCT, Portugal).

\end{document}


\title{Supplementary material}




%


%

\maketitle

\section{The $Z=1, 2$ isotopes H, He for different superheavy systems showing ternary fission}
\label{App:1}

As shown for the ternary fission of $^{252}$Cf(sf) in \cite{rnp20}, we give also the corresponding calculations for the other  ternary fission systems, 
see Tabs. \ref{Tab:U233}, \ref{Tab:U235}, \ref{Tab:Pu239}, \ref{Tab:Pu241}, \ref{Tab:Cm248}, \ref{Tab:Cf252}. 
Data $Y^{\rm obs}_{A,Z}$ are taken from \cite{Koester,Valskii2,Kopatch02,KoesterPhD}.\\

\begin{table}
\begin{center}
\hspace{0.5cm}
 \begin{tabular}{|c|c|c|c|c|c|c|c|c|c|c|c|c|c|c|}
\hline
isotope& $A$  &  $Z$ & $Y^{\rm obs}_{A,Z}$& $Y^{\rm interp}_{A,Z}$&  $\frac{B_{A,Z}}{A} $& $ g_{A,Z} $ & $Y^{\rm final}_{A,Z}$ & $E^{\rm thresh}_{A,Z}$  &$R^\gamma_{A,Z}(1.3)$& $Y^{\rm rel,\gamma}_{A,Z}$ 
& $R_{A,Z}^{\rm vir}(1.3)$ & $Y^{\rm rel, vir}_{A,Z}$ & $R_{A,Z}^{\rm eff}(1.3)$ & $Y^{\rm rel, eff}_{A,Z}$\\
\hline
 $\lambda_T$ 	& -&-&-&(1.17)	&-&-		 &1.23337 &-&-& 1.25999 &- &1.24232 & - & 1.24836  \\
$\lambda_n$ 	& -&-&-&(-3.2)	&-&-		 &-3.53462 &-&-&-3.56867 &- &-3.52192 & - &-3.51185 \\
$\lambda_p$ 	& -&-&-&(-16.0)&-&-		 &-15.6858 &-&-&-15.7094 &- &-15.8192 & - &-15.8852 \\
\hline
$^1$n		&1& 0	&-			&2.13e6 	&0		& 2 &533268	        &-              &-              & 428063        	&-              &557240  &-              & 5578702\\
$^1$H		&1& 1 	& -			& 38		&0		& 2 &28.072 	        &-              &-              &27.9787         	&-              &27.998 &-              	& 27.665 \\
$^2$H 		&2& 1 	& 41(2) 		& 48		& 1.112 	& 3 &41.154		&2.224	&1		&40.8301		& 0.98	&40.95  & 0.98		& 41 \\
$^3$H$^{\rm obs}$&3& 1 	& 460(20)		& 630	& 2.827 	& 2 & 458.21.		&-		&-		&461.886		& -		&460.56 & -		&460 \\
$^3$H		&3& 1 	& -			&-		& 2.827 	& 2 & {\it (458.21)}	&6.257	&1		&422.322		& 0.99	&457.98 & 0.99		&457.36 \\
$^4$H		&4& 1 	& -			& -		& 1.720	& 5 & {\it  (27.419 )}	&-1.6	&1.473	&39.5642		&0.0606	&2.5741 & 0.0606	&2.6437 \\
$^3$He		&3& 2 	& $< 0.01$	&  0.0058	& 2.573	& 2 &0.012972		&5.493	&1		&0.01504 		& 0.988	& 0.01243 & 0.988	&0.012265 \\
$^4$He$^{\rm obs}$	&4&2 & 10000		& 10000	& 7.073 	& 1 & 10000  		&-		&- 		& 10000  		& - 		&10000   & - 		&10000\\
$^4$He		&4& 2 	& -			& -		& 7.073 	& 1 &  {\it (10000)}  	&20.577	&1 		&8429.0	 	& 1 		&8858.5 & 1 		&8255.9 \\
$^5$He		&5& 2 	& -			&1500	& 5.512	& 4 &  {\it (1758.2)} 	&-0.735	&1		&1551.8  		&0.7044 	&1135.6 & 1.12511 	&1733.7 \\
$^6$He$^{\rm obs}$	&6&2&137(7)		& 140	& 4.878 	& 1 & 132.04		& -		&-		&142.05  		& - 		&138.4 & - 		&137.0  \\
$^6$He		&6& 2 	& -			& -		& 4.878 	& 1 &   {\it (132.04)}  & 0.975	&1		&117.03 		& 0.9453 	&116.86 &0.943829 	&113.22 \\
$^7$He		&7& 2 	& -			& 28		& 4.123	& 4 & {\it (27.111) } 	&-0.410 	&1		&25.024  		& 0.821 	& 21.544 &0.912131 &23.777 \\
$^8$He$^{\rm obs}$	&8&2& 3.6(4)		& 4.6		& 3.925 	& 1 & 3.67011		& -		&-		&3.5299		&- 		&3.5797 &- 		&3.6 \\
$^8$He		&8& 2 	& -			& - 		& 3.925 	& 1 &  {\it (3.6701)}	& 2.125	&1		&3.3544 		&0.9783 	& 3.5316 &0.9783 	&3.5503 \\
$^9$He		&9& 2 	& -			& - 		& 3.349	& 2 &  {\it (0.18085)}  & -1.25	&1		&0.17463   	& 0.2604 	&0.04812 & 0.2604  	&0.04968 \\
$^8$Be		&8& 4	& 0.5(2)		& 2.8		& 7.062 	& 1 &  {\it (7.0298)} &-0.088 	&1.49 	&9.6012     	 &1.07        &5.8212 &1.07        	&5.17019 \\
\hline
fit metric		&-&-		 & -	&-		&-		&-   & 0.0005874	 & - 		& - 		&0.000435 	& - 		&0.0000346 	& - 	& 0\\
\hline
 \end{tabular}
\caption{
Properties and yields of the H, He and Be isotopes from ternary fission of $^{233}$U($n_{\rm th}$,f) relevant for the final distribution of observed H, He nuclei (denoted by the superscript 'obs'). 
Experimental  yields $Y^{\rm obs}_{A,Z}$  are compared to the yields $Y^{\rm interp}_{A,Z}$ obtained from an interpolation formula \cite{Valskii2} as well the yields calculated 
in different approximations of the nuclear Hamiltonian $H$, see \cite{rnp20}, together with the corresponding multipliers  $R_{A,Z}(\lambda_T)$, which represent the intrinsic partition functions:  
The final state distribution $Y^{\rm final}_{A,Z}$ is constructed considering only the observed stable nuclei $\{A,Z\}^{\rm obs}$. The distribution (3) of Ref. \cite{rnp21} is used to infer the 
optimum choices for the Lagrange parameters $\lambda^{\rm final}_i$, with the prefactor $R^{\rm final}_{A,Z}=1$. The other yields of unstable nuclei are calculated with these  Lagrange parameters and are shown in parentheses.
The relevant distribution $Y^{\rm rel,\gamma}_{A,Z}$ of noninteracting  nuclei (including excited bound states performing internal transitions with emission of $\gamma$ rays) is constructed including all unstable states of nuclei, 
the corresponding intrinsic partition function of the channel $\{A,Z\}$ is denoted  by $R^\gamma_{A,Z}(\lambda_T)$, and the final distribution follows as sum of all feeding 
clusters. The Lagrange parameters $\lambda^\gamma_i$ are obtained optimizing these final distribution versus the observed distribution $Y^{\rm obs}_{A,Z}$.
The relevant distribution $Y^{\rm rel,vir}_{A,Z}$ takes into account continuum correlations, and the relevant distribution $Y^{\rm rel, eff}_{A,Z}$ including interaction between the constituents.  
 The binding energy $B_{A,Z}$ (in MeV), the degeneracy $g_{A,Z}$, and  threshold energy of continuum states $E^{\rm thresh}_{A,Z}$ (in MeV) according Ref. \cite{nuclei}, are also given. 
(Note that the threshold of continuum states for $^6$He, $^8$He is given by the emission of two neutrons $S_{2n}$.)
 The first three rows show the Lagrange parameters $\lambda_i$ obtained for the four different calculations (in  MeV). For the column $Y^{\rm interp}_{A,Z}$, instead of the Lagrange parameters the corresponding parameter values for temperature and chemical potentials according to \cite{Valskii2} are given (in parentheses).}
\label{Tab:U233}
\end{center}
\end{table}

\begin{table}
\begin{center}
\hspace{0.5cm}
 \begin{tabular}{|c|c|c|c|c|c|c|c|c|c|c|c|c|c|c|}
\hline
isotope& $A$  &  $Z$ & $Y^{\rm obs}_{A,Z}$& $Y^{\rm interp}_{A,Z}$&  $\frac{B_{A,Z}}{A} $& $ g_{A,Z} $ & $Y^{\rm final}_{A,Z}$ & $E^{\rm thresh}_{A,Z}$  &$R^\gamma_{A,Z}(1.3)$& $Y^{\rm rel,\gamma}_{A,Z}$ 
& $R_{A,Z}^{\rm vir}(1.3)$ & $Y^{\rm rel, vir}_{A,Z}$ & $R_{A,Z}^{\rm eff}(1.3)$ & $Y^{\rm rel, eff}_{A,Z}$\\
\hline
 $\lambda_T$ 	& -&-&-&(1.31)	&-&-		 & 1.21606 &-&-& 1.24608 &- &1.22411 & - & 1.23613  \\
$\lambda_n$ 		& -&-&-&(-3.3)	&-&-		 &-3.2548 &-&-&-3.2872 &- &-3.23611 & - &-3.23168 \\
$\lambda_p$ 		& -&-&-&(-17.1)&-&-		 &-16.2959 &-&-&-16.3407 &- & -16.458 & - &-16.4754 \\
\hline
$^1$n		&1& 0	&-			&5.99e6 	&0		& 2 &1.24722e6        &-               &-               &968123        &-              &1.3405e6  &-              & 1.22206e6\\
$^1$H		&1& 1 	& 115(15)		& 160	&0		& 2 &27.4494 	        	&-               &-               &27.3158         	&-              	& 27.2132 	&-              &27.173 \\
$^2$H 		&2& 1 	& 50(2) 		& 204	& 1.112 	& 3 & 49.8926		&2.224	&1		&49.3736		& 0.98	&49.495 	& 0.98	& 50 \\
$^3$H$^{\rm obs}$&3& 1 	& 720(30)		& 1920	& 2.827 	& 2 & 721.55		&-		&-		&729.068		& -		& 727.234 	& -		& 720 \\
$^3$H		&3& 1 	& -			&-		& 2.827 	& 2 & {\it (721.55)}	&6.257	&1		&655.536		& 0.99	&722.411 	& 0.99	&715.021 \\
$^4$H		&4& 1 	& -			& -		& 1.720	& 5 & {\it  (51.2211 )}	&-1.6	&1.473	&73.5314		&0.0606	& 4.82309  	& 0.0606	&4.97869 \\
$^3$He		&3& 2 	& $< 0.01$	&  0.0287	& 2.573	& 2 &0.00846543	&5.493	&1		&0.0100105 	& 0.988	& 0.00785028	& 0.988	& 0.00854506 \\
$^4$He$^{\rm obs}$	&4&2 & 10000		& 10020	& 7.073 	& 1 & 10000  		&-		&- 		& 10000  		& - 		&10000  	& - 		&10000\\
$^4$He		&4& 2 	& -			& -		& 7.073 	& 1 &  {\it (10000)}  	&20.577	&1 		&8174.46  	& 1 		&8660.91 	& 1 		& 8259.7 \\
$^5$He		&5& 2 	& -			& 1980	& 5.512	& 4 &  {\it (2106.72)} 	&-0.735	&1		&1815.85  	&0.7044 	&1333.49 	& 0.927608 	&1734.54 \\
$^6$He$^{\rm obs}$	&6&2&191(9)		& 193	& 4.878 	& 1 & 195.006  		& -		&-		&212.531 		& - 		& 207.271	 & - 		&191  \\
$^6$He		&6& 2 	& -			& -		& 4.878 	& 1 &   {\it (195.006)}  & 0.975	&1		&168.845  	& 0.9453 	& 169.599  	&0.876854 	&157.851 \\
$^7$He		&7& 2 	& -			& 50.0	& 4.123	& 4 &  {\it (48.1545) } &-0.410 	&1		&43.6861 		& 0.821 	& 37.672 	&0.696886 	&33.1488 \\
$^8$He$^{\rm obs}$	&8&2& 8.2(6)		& 8.0		& 3.925 	& 1 & 8.11096 		& -		&-		&7.72776 		&- 		&7.83298 	&- 		& 8.2 \\
$^8$He		&8& 2 	& -			& - 		& 3.925 	& 1 &  {\it (8.11096)}	& 2.125	&1		&7.273		&0.9783 	& 7.7077 	&0.9783 	&8.06369 \\
$^9$He		&9& 2 	& -			& - 		& 3.349	& 2 &  {\it (0.47606)}  & -1.25	&1		&0.45476	 	& 0.2604 	&0.125284 	& 0.2604  &0.136306 \\
$^8$Be		&8& 4	& 0.5(2)		& 1.2		& 7.062 	& 1 &  {\it (3.62842) } &-0.088 	&1.49 	&4.84514         	&1.07        &2.80146 	&1.07        &2.87999 \\
\hline
fit metric		&-&-		 & -	&-		&-		&-   & 0.000139849	 & - 		& - 		&0.00381071 	& - 		&0.00224614 	& - 	& 0\\
\hline
 \end{tabular}
\caption{
Properties and yields of the H, He and Be isotopes  from ternary neutron-induced fission $^{235}$U($n_{\rm th}$,f), see Tab. \ref{Tab:U233}.}
\label{Tab:U235}
\end{center}
\end{table}

\begin{table}
\begin{center}
\hspace{0.5cm}
 \begin{tabular}{|c|c|c|c|c|c|c|c|c|c|c|c|c|c|c|}
\hline
isotope& $A$  &  $Z$ & $Y^{\rm obs}_{A,Z}$& $Y^{\rm interp}_{A,Z}$&  $\frac{B_{A,Z}}{A} $& $ g_{A,Z} $ & $Y^{\rm final}_{A,Z}$ & $E^{\rm thresh}_{A,Z}$  &$R^\gamma_{A,Z}(1.3)$& $Y^{\rm rel,\gamma}_{A,Z}$ 
& $R_{A,Z}^{\rm vir}(1.3)$ & $Y^{\rm rel, vir}_{A,Z}$ & $R_{A,Z}^{\rm eff}(1.3)$ & $Y^{\rm rel, eff}_{A,Z}$\\
\hline
 $\lambda_T$ 	& -&-&-&(1.37)		&-&-		 &1.30982 &-&-&1.35068 &- &1.31915 & - & 1.33895  \\
$\lambda_n$ 		& -&-&-&(-3.4)	&-&-		 &-3.43154 &-&-&-3.48007&- &-3.40999 & - &-3.41297 \\
$\lambda_p$ 		& -&-&-&(-16.8)&-&-		 &-16.0317 &-&-&-16.0629 &- &-16.2294 & - &-16.1973\\
\hline
$^1$n		&1& 0	&-			&2.8e6 	&0		& 2 &611234	          &-               &-               & 449548       &-              &666112   	&-              & 565349\\
$^1$H 		&1& 1 	&190(10)		& 164	&0		& 2 &40.5879	        	&-               &-               & 40.4499         &-              	&40.0952 	&-              & 40.3306 \\
$^2$H		&2& 1 	& 69(2) 		& 205	& 1.112 	& 3 &68.4939		&2.224	&1		&67.7145		& 0.98	&67.8421 	& 0.98	&69.0\\
$^3$H$^{\rm obs}$&3& 1 	& 720(30)		& 1620	& 2.827 	& 2 & 725.32		&-		&-		&733.446		& -		&732.186 	& -		&720.0 \\
$^3$H		&3& 1 	& -			&-		& 2.827 	& 2 & {\it 725.32)}	&6.257	&1		&648.055		& 0.99	&726.534 	& 0.99	&714.137 \\
$^4$H		&4& 1 	& -			& -		& 1.720	& 5 & {\it  (59.8756)}	&-1.6	&1.473	&85.391		&0.0606	& 5.65192  	& 0.0606	&5.86338 \\
$^3$He		&3& 2 	& $< 0.01$	&  0.0055	& 2.573	& 2 &0.0268577	&5.493	&1		&0.0330962	& 0.988	&0.0244876	& 0.988	&0.0287137 \\
$^4$He$^{\rm obs}$	&4&2 & 10000		& 9900	& 7.073 	& 1 & 10000  		&-		&- 		& 10000  		& - 		&10000  	& - 		&10000\\
$^4$He		&4& 2 	& -			& -		& 7.073 	& 1 &  {\it (10000)}  	&20.577	&1 		&8000.67  	& 1 		& 8533.71 	& 1 		&8253.43 \\
$^5$He		&5& 2 	& -			&2100	& 5.512	& 4 &  {\it (2327.71)} 	&-0.735	&1		&1977.95  	&0.7044 	&1454.62 	& 0.830305 	&1733.22 \\
$^6$He$^{\rm obs}$	&6&2&192(5)		&205	& 4.878 	& 1 &206.139		& -		&-		&225.456  	& - 		&220.159	 & - 		&192  \\
$^6$He		&6& 2 	& -			& -		& 4.878 	& 1 &   {\it (206.139)} & 0.975	&1		&175.821		& 0.9453 	& 177.356 &0.822729 	&158.678 \\
$^7$He		&7& 2 	& -			& 56.5	& 4.123	& 4 &  {\it (55.197) } 	&-0.410 	&1		&49.6348  	& 0.821 	&42.8027 &0.596994 &33.3223 \\
$^8$He$^{\rm obs}$	&8&2&8.8(4)		&8.7		& 3.925 	& 1 & 8.47722	 	& -		&-		& 8.04813		&- 		&8.14964 &- 		&8.8 \\
$^8$He		&8& 2 	& -			& - 		& 3.925 	& 1 &  {\it (8.47722)}	& 2.125	&1		& 7.5085		&0.9783 	&8.00119 &0.9783 	&8.63165 \\
$^9$He		&9& 2 	& -			& - 		& 3.349	& 2 &  {\it (0.566804)} & -1.25	&1		&0.539634  	& 0.2604 	&0.1484488 &0.26959&0.168347 \\
$^8$Be		&8& 4	& -			& 2.7		& 7.062 	& 1 &  {\it (7.8761) } 	&-0.088 	&1.49 	& 10.6877         	&1.07        &5.83416	 &1.07        &6.67211 \\
\hline
fit metric		&-&-		 & -	&-		&-		&-   & 0.00163846	 & - 		& - 		&0.00861863 	& - 		&0.00629828 	& - 	& 0\\
\hline
 \end{tabular}
\caption{
Properties and yields of the H, He and Be isotopes  from ternary  neutron-induced  fission $^{239}$Pu($n_{\rm th}$,f), see Tab. \ref{Tab:U233}.}
\label{Tab:Pu239}
\end{center}
\end{table}

\begin{table}
\begin{center}
\hspace{0.5cm}
 \begin{tabular}{|c|c|c|c|c|c|c|c|c|c|c|c|c|c|c|}
\hline
isotope& $A$  &  $Z$ & $Y^{\rm obs}_{A,Z}$& $Y^{\rm interp}_{A,Z}$&  $\frac{B_{A,Z}}{A} $& $ g_{A,Z} $ & $Y^{\rm final}_{A,Z}$ & $E^{\rm thresh}_{A,Z}$  &$R^\gamma_{AZ}(1.3)$& $Y^{\rm rel,\gamma}_{A,Z}$ 
& $R_{A,Z}^{\rm vir}(1.2)$ & $Y^{\rm rel, vir}_{A,Z}$ & $R_{A,Z}^{\rm eff}(1.2)$ & $Y^{\rm rel, eff}_{A,Z}$\\
\hline
 $\lambda_\beta$ 	& -&-&-	&(1.38)	&-&-	& 1.18802 &-&-		& 1.21997 &- 	& 1.19574 & - 	&1.206  \\
$\lambda_n$ 		& -&-&-	&(-3.2)	&-&-	& -3.0161 &-&-		&-3.04481 &- 	&-2.99366 & - 	&-2.9903 \\
$\lambda_p$ 		& -&-&-	&(-17.2)&-&-	& -16.4794 &-&-		&-16.5357 &-	&-16.6651 & - 	&-16.6615 \\
\hline
$^1$n		&1& 0	&-	&4.3e6 	&0		& 2 & 1.60897e6        &-      	&-      & 1.21609e6      	 &-      	&1.75799e6   	&-          	&1.59781e6 \\
$^1$H		&1& 1 	& -	& 162	&0		& 2 & 19.2691 	        	&-      	&-      & 19.1596        	&-      	& 19.0344    		 &-          	&19.0696 \\
$^2$H 		&2& 1 	& 42 	& 240	& 1.112 	& 3 & 41.9699		&2.224	&1	& 41.4784		& 0.98	& 41.5776    	& 0.98	& 42 \\
$^3$H$^{\rm obs}$&3&1 	& 786& 2230	& 2.827 	& 2 & 786.564		&-		&-	&795.847		& -		&793.931   	& -		& 786 \\
$^3$H		&3& 1 	& -	&-		& 2.827 	& 2 & {\it (786.564)}	&6.257	&1	& 706.914		& 0.99	&788.063   	& 0.99	& 779.982 \\
$^4$H		&4& 1 	& -	& -		& 1.720	& 5 & {\it  (62.1225)}	&-1.6	&1.473& 88.9333	&0.0606	&5.86747		 & 0.0606	&6.01689 \\
$^3$He		&3& 2 	& -	&  -		& 2.573	& 2 & 0.00494758	&5.493	&1	&0.005949145 	& 0.988	&0.00450025  	& 0.988	&0.00492648 \\
$^4$He$^{\rm obs}$&4&2 & 10000	& 9.81e3	& 7.073 & 1 & 10000  	&-		&- 	& 10000  		& - 		&10000     	 & - 		&10000\\
$^4$He		&4& 2 	& -	& -		& 7.073 	& 1 & {\it (10000)}  	&20.577	&1 	&7974.82  	& 1 		& 8509.38  	 & 1 			&8260.3 \\
$^5$He		&5& 2 	& -	& 2.4e3	& 5.512	& 4 & {\it (2383.77)} 	&-0.735	&1	& 2016.72  	&0.70279&1485.88     	& 0.822704 	&1734.66 \\
$^6$He$^{\rm obs}$&6&2 	& 260	& 288	& 4.878 & 1 &261.806  		& -	&-& 288.58 	 & - 		& 281.697   	& - 			& 260  \\
$^6$He		&6& 2 	&-	& -		& 4.878 	& 1 & {\it (261.806)}  	& 0.975	&1	& 222.641  	& 0.9453	&224.733    	&0.893547 	&214.876 \\
$^7$He		&7& 2 	& -	& -		& 4.123	& 4 & {\it (73.6062)} 	&-0.410 	&1	& 65.9395  	& 0.821 	& 56.9648  	 &0.625792 	&45.124 \\
$^8$He$^{\rm obs}$&8&2 	& 15		& 17.9	& 3.925 & 1 & 14.9455 		& -	&-& 14.1504 	 &- 		& 14.338  	&- 			&15 \\
$^8$He		&8& 2 	& -	& - 		& 3.925 	& 1 & {\it (14.9455)}	& 2.125	&1	&13.2179	 	&0.9783 	&14.081    	&0.9783 		& 14.7224 \\
$^9$He		&9& 2 	& -	& - 		& 3.349	& 2 & {\it (0.982611)}  	& -1.25	&1	&0.932532   	& 0.2604	&0.256979 	 & 0.2604  	&0.277667 \\
$^8$Be		&8& 4	& -	& 2.1		& 7.062 	& 1 & {\it (3.22254)} 	&-0.088 	&1.49 	&4.22574       &1.07   & 2.36803  	 &1.07       		&2.51651 \\
\hline
fit metric		&-&-		 & -	&-		&-		&-   & 0.0000207417	 & - 		& - 		&0.0036469 	& - 		&0.00216637 	& - 	& 0\\
\hline
 \end{tabular}
\caption{
Properties and yields of the H, He and Be isotopes  from ternary fission $^{241}$Pu($n_{\rm th}$,f), see Tab. \ref{Tab:U233}.} 
\label{Tab:Pu241}
\end{center}
\end{table}

\begin{table}
\begin{center}
\hspace{0.5cm}
 \begin{tabular}{|c|c|c|c|c|c|c|c|c|c|c|c|c|c|c|}
\hline
isotope& $A$  &  $Z$ & $Y^{\rm obs}_{A,Z}$& $Y^{\rm interp}_{A,Z}$&  $\frac{B_{A,Z}}{A} $& $ g_{A,Z} $ & $Y^{\rm final}_{A,Z}$ & $E^{\rm thresh}_{A,Z}$  &$R^\gamma_{A,Z}(1.3)$& $Y^{\rm rel,\gamma}_{A,Z}$ 
& $R_{A,Z}^{\rm vir}(1.3)$ & $Y^{\rm rel, vir}_{A,Z}$ & $R_{A,Z}^{\rm eff}(1.3)$ & $Y^{\rm rel, eff}_{A,Z}$\\
\hline
 $\lambda_T$ 		& -&-&-&(1.20)	&-&-		 &1.22736 &-&-& 1.26954 &- &1.23655 & - &1.24232 \\
$\lambda_n$ 		& -&-&-&(-2.3)	&-&-		 &-2.93864 &-&-&-2.97442 &- &-2.91161 & - &-2.91205 \\
$\lambda_p$ 		& -&-&-&(-16.5)&-&-		 &-16.5591 &-&-&-16.6174 &-&-16.7822 & - &-16.7633 \\
\hline
$^1$n		&1& 0	&-			&2.3e6 	&0		& 2 &1.39802e6	&-               &-               & 980306        &-              &1.55008e6  &-              & 1.45473e6\\
$^1$H		&1& 1 	& 160(20)		& 26.6	&0		& 2 &21.1822	        	&-               &-               & 21.0979         &-              	& 20.8366 &-              &20.9234 \\
$^2$H 		&2& 1 	& 50(5) 		& 51		& 1.112 	& 3 &50.2036		&2.224	&1		&49.5638		& 0.98	&49.681 	& 0.98	& 50 \\
$^3$H$^{\rm obs}$&3& 1 	& 922(18)		& 921	& 2.827 	& 2 & 918.26		&-		&-		&929.956		& -		&927.894 	& -		&922 \\
$^3$H		&3& 1 	& -			&-		& 2.827 	& 2 & {\it (918.26)}	&6.257	&1		& 805.649		& 0.99	&919.588 	& 0.99	&913.601 \\
$^4$H		&4& 1 	& -			& -		& 1.720	& 5 & {\it  (87.4953 )}	&-1.6	&1.473	&124.307		&0.0606	&8.30575  & 0.0606	&8.38967 \\
$^3$He		&3& 2 	& -			&  -		& 2.573	& 2 &0.00745986	&5.493	&1		&0.00949126 	& 0.988	&0.00665866& 0.988	&0.007116 \\
$^4$He$^{\rm obs}$	&4&2 & 10000		& 10100	& 7.073 	& 1 & 10000  		&-		&- 		& 10000  		& - 		&10000  	& - 		&10000\\
$^4$He		&4& 2 	& -			& -		& 7.073 	& 1 &  {\it (10000)}  	&20.577	&1 		&7673.86  	& 1 		&8281.42 	& 1 		& 8259.22 \\
$^5$He		&5& 2 	& -			& -		& 5.512	& 4 &  {\it (2809.23)} 	&-0.735	&1		& 2314.8	  	&0.7044 	&1712.66	& 0.705818 	& 1734.44 \\
$^6$He$^{\rm obs}$	&6&2&354(31)		& 335	& 4.878 	& 1 & 340.382		& -		&-		&380.437		& - 		&372.085	 & - 		& 354  \\
$^6$He		&6& 2 	& -			& -		& 4.878 	& 1 &   {\it (340.382)} & 0.975	&1		&281.87 		& 0.9453 	& 286.755 &0.950295 & 292.562 \\
$^7$He		&7& 2 	& -			& -		& 4.123	& 4 &  {\it (111.811) } &-0.410 	&1		& 98.5665		& 0.821 	& 85.3308 &0.575412 & 61.438 \\
$^8$He$^{\rm obs}$	&8&2&24(4)		& 26		& 3.925 	& 1 & 24.5003	 	& -		&-		& 23.0475		&- 		&23.3302 &- 		&24 \\
$^8$He		&8& 2 	& -			& - 		& 3.925 	& 1 &  {\it (24.5003)}	& 2.125	&1		& 21.2309		&0.9783 	& 22.8296 &0.9783 	&23.4773 \\
$^9$He		&9& 2 	& -			& - 		& 3.349	& 2 &  {\it (1.92507)}  & -1.25	&1		& 1.81659  	& 0.2604 	&0.500566 & 0.2337 &0.52265 \\
$^8$Be		&8& 4	&-			& 4.0		& 7.062 	& 1 &  {\it (4.29545) } &-0.088 	&1.49 	& 5.67174       	&1.07        &2.95952 &1.07        & 3.17102 \\
\hline
fit metric		&-&-		 & -			&-		&-		&-   & 0.000665883	 & - 		& - 		&0.00174446 	& - 		& 0.000841528 	& - 	& 0\\
\hline
 \end{tabular}
\caption{
Properties and yields of the H, He and Be isotopes  from ternary  spontaneous fission $^{248}$Cm(sf), see Tab. \ref{Tab:U233}.} 
\label{Tab:Cm248}
\end{center}
\end{table}

\begin{table}
\begin{center}
\hspace{0.5cm}
 \begin{tabular}{|c|c|c|c|c|c|c|c|c|c|c|c|c|c|c|}
\hline
isotope& $A$  &  $Z$ & $Y^{\rm obs}_{A,Z}$& $Y^{\rm interp}_{A,Z}$&  $\frac{B_{A,Z}}{A} $& $ g_{A,Z} $ & $Y^{\rm final}_{A,Z}$ & $E^{\rm thresh}_{A,Z}$  &$R^\gamma_{A,Z}(1.3)$& $Y^{\rm rel,\gamma}_{A,Z}$ 
& $R_{A,Z}^{\rm vir}(1.3)$ & $Y^{\rm rel, vir}_{A,Z}$ & $R_{A,Z}^{\rm eff}(1.3)$ & $Y^{\rm rel, eff}_{A,Z}$\\
\hline
 $\lambda_T$ 	& -&-&-&(1.25)	&-&-		 & 1.26406 &-&-& 1.30761 &- &1.2704 & - & 1.30501  \\
$\lambda_n$ 		& -&-&-&(-2.8)	&-&-		 &-3.03054 &-&-& --3.07019 &- & -2.9993 & - &-3.0071 \\
$\lambda_p$ 		& -&-&-&(-15.8)&-&-		 & -16.5313 &-&-&-16.5961 &- & -16.7773 & - &-16.6477 \\
\hline
$^1$n		&1& 0	&-			&0.65e6 	&0		& 2 &1.19729e6        &-               &-               & 848881        &-              &1.37951e6  	&-              & 956227.\\
$^1$H		&1& 1 	& 160(20)		& 19.4	&0		& 2 & 27.5243	        	&-               &-               & 27.3234        &-              & 26.8909 		&-              & 27.6113 \\
$^2$H 		&2& 1 	& 63(3) 		& 37.5	& 1.112 	& 3 & 61.6948		&2.224	&1		& 60.6916		& 0.98	& 60.7334 	& 0.98	&63. \\
$^3$H$^{\rm obs}$	&3& 1 	& 950(90)	& 591	& 2.827 	& 2 & 970.097		&-		&-		&985.789		& -		& 985.294 	& -		&950. \\
$^3$H		&3& 1 	& -			&-		& 2.827 	& 2 & {\it (970.097)}	&6.257	&1		&850.357		& 0.99	& 976.224 	& 0.99	& 940.44 \\
$^4$H		&4& 1 	& -			& -		& 1.720	& 5 & {\it  (95.6973)}	&-1.6	&1.473	&135.432		&0.0606	& 9.06948  	& 0.0606	& 9.56035 \\
$^3$He		&3& 2 	& -			&  0.0095	& 2.573	& 2 &0.0121759	&5.493	&1		&0.015248	& 0.988	& 0.0104211	& 0.988	& 0.0150797 \\
$^4$He$^{\rm obs}$	&4& 2 	& 10000		& 10028	& 7.073 	& 1 & 10000  	&-		&- 		& 10000  		& - 		&10000  		& - 		&10000\\
$^4$He		&4& 2 	& 8264(341)	& -		& 7.073 	& 1 &  {\it (10000)}  	&20.577	&1 		&7651.6  		& 1 		& 8267.37 	& 1 		& 8256.14\\
$^5$He		&5& 2 	& 1736(274)	& 2600	& 5.512	& 4 &  {\it (2849.21)} 	&-0.735	&1		&2335.42  	&0.7044 	&1726.03 		& 0.65937 &1733.79 \\
$^6$He$^{\rm obs}$	&6& 2 	& 270(30)		& 369	& 4.878 	& 1 & 330.455  	& -		&-		& 367.607  	& - 		&359.343	 	& - 		& 270.  \\
$^6$He		&6& 2 	& 223(26)		& -		& 4.878 	& 1 &   {\it (330.455)}  & 0.975	&1		&272.05  		& 0.9453 	& 276.769  	&0.695557 & 223.14 \\
$^7$He		&7& 2 	& 47(9)		& 128	& 4.123	& 4 &  {\it (109.268) } &-0.410 	&1		&95.5576  	& 0.821 	&82.5742 		&0.398335 & 46.8595 \\
$^8$He$^{\rm obs}$	&8& 2 	& 25(5)		& 30.6	& 3.925 	& 1 & 22.4798 	& -		&-		& 21.0178  	&- 		& 21.2537 	&- 		& 25. \\
$^8$He		&8& 2 	& 25(5)		& - 		& 3.925 	& 1 &  {\it (22.4798)}	& 2.125	&1		& 19.3247 	&0.9783 	& 20.7884 	&0.9783 	& 24.4066 \\
$^9$He		&9& 2 	& -			& - 		& 3.349	& 2 &  {\it (1.81358)}  & -1.25	&1		& 1.69311   	& 0.2604 	&0.465325 	& 0.2604  &0.593398 \\
$^8$Be		&8& 4	& 10(6)		& 15.2	& 7.062 	& 1 &  {\it (5.01011) } &-0.088 	&1.49 	& 6.49241         &1.07        & 3.29964 	&1.07        & 5.03318 \\
\hline
fit metric		&-&-		 & -			&-		&-		&-   & 0.01766            & -             & -             & 0.03202          & -             & 0.02768          & -             & 0\\
\hline
 \end{tabular}
\caption{
Properties and yields of the H, He and Be isotopes  from ternary  spontaneous fission $^{252}$Cf(sf), see Tab. \ref{Tab:U233} and \cite{rnp20}.}
\label{Tab:Cf252}
\end{center}
\end{table}

\vspace{3cm}
As in Ref. \cite{Sara}, we use in this section for the fit metric that of Lestone \cite{Lestone08},
defined by
\begin{equation}
M^2=\sum_{A,Z}^n \left(\ln[Y_{A,Z}^{\rm final,x}]-\ln[Y_{A,Z}^{\rm obs}]\right)^2/n
\end{equation}
where $n$ is the number of fitted experimental data
points. The exponential of $M$ is a measure of the typical
relative difference between the model calculations and the
experimental data. For $M \approx 1$ the average relative discrepancy
between model and experiment would be a factor of $\approx 3$. 
In our other calculations, the minimum of the expression $\sum_{A,Z}^n \left(Y_{A,Z}^{\rm final,x}-Y_{A,Z}^{\rm obs}\right)^2/Y_{A,Z}^{\rm obs}$ has been determined which gives approximately the same fit results.

The fit $Y_{A,Z}^{\rm final}$ is performed for observed isotopes $^2$H, $^3$H, $^6$He, $^8$He; the value for $^4$He is fixed by normalization.
They can be represented by a statistical distribution with the corresponding Lagrange parameters $\lambda_i$. However, also other isotopes are expected
from this statistical distribution (NSE), and the corresponding yields are given in parentheses. If we take these into account, inclusive excited states which decay into the ground state, and unstable nuclei which feed stable ones, we arrive at $Y_{A,Z}^{\rm rel, \gamma}$. The fit of the corresponding final yields which are calculation as sum of the feeding primary yields to the observed yields is characterized by the corresponding fit metric. The account of continuum correlations gives the prefactor $R_{A,Z}^{\rm vir}(\lambda_T)$ which is related to the intrinsic partition function of the corresponding channel. We use the known expressions for the virial coefficients in some channels as calculated from the Beth-Uhlenbeck formula, values are taken from \cite{rnp20}. Estimates can also be given according  the relation given in \cite{rnp20}, see Eq. (\ref{fitRvir}) below, which gives the values $R_{^2{\rm H}}^{\rm vir}(1.3)=0.973$, $R_{^3{\rm H}}^{\rm vir}=0.998$, $R_{^4{\rm H}}^{\rm vir}=0.0652$, $R_{^3{\rm He}}^{\rm vir}=0.997$, $R_{^4{\rm He}}^{\rm vir}=1$, $R_{^5{\rm He}}^{\rm vir}=0.689$, $R_{^6{\rm He}}^{\rm vir}=0.933$, $R_{^7{\rm He}}^{\rm vir}=0.875$, $R_{^8{\rm He}}^{\rm vir}=0.971$, $R_{^9{\rm He}}^{\rm vir}=0.255$, these values have been used in Tabs. \ref{all241Pu} - \ref{all245Cm} below. In principle, the parameter $\lambda_T$ should be determined self-consistently, but we use here a representative value $\lambda_T=1.3$ MeV.

We see that the fit metric is improved going from $Y^{\rm rel,\gamma}_{A,Z}$ to $Y^{\rm rel, vir}_{A,Z}$. This may be considered as a proof that continuum states have to be included, and the need of a virial expansion. The good results for the fit metric of $Y^{\rm final}_{A,Z}$ is accidental. It is clear that also the unstable isotopes are produced in the scission process and have been identified, but they are not as important as assumed within the NSE.

The yields $Y_{A,Z}^{\rm rel, eff}$ have been constructed in the following way: There are deviations between the final yields obtained from the primary yields 
$Y^{\rm rel, vir}_{A,Z}$ and the observed yields. One possible effect is Pauli blocking as an in-medium effect, which suppresses in particular weakly bound states such as continuum correlations ($^4$H, $^5$He, $^7$He, $^9$He), but also $^6$He which is weakly bound. The primary yields of $^5$He, $^7$He
have been observed \cite{Kopatch02} for $^{252}$Cf and are taken here to determine  effective values $R_{^5{\rm He}}^{\rm eff}$,  $R_{^7{\rm He}}^{\rm eff}$ which are needed to reproduce them.  Because data for  primary yields of $^5$He, $^7$He are not available for other fissioning actinides,
the observed fractions measured for $^{252}$Cf have been used for all actinides to derive  effective values $R_{^5{\rm He}}^{\rm eff}$,  $R_{^7{\rm He}}^{\rm eff}$. This assumption is not well justified, and these values will be improved when measurements for the other actinides are available. The same holds also for $^8$Be where measurements have been performed for  $^{252}$Cf only \cite{Jesinger05}.

Furthermore, the observed 
yield for  $^6$He was used to determine the corresponding effective  value $R_{^6{\rm He}}^{\rm eff}(1.3)$. The other isotopes are stronger bound so that the influence of medium effects is supposed to be small. The value of $R_{^6{\rm He}}^{\rm eff}(1.3)$ was considered individually for each fissioning actinide.

\section{Other isotopes observed in the ternary fission of $^{241}$Pu($n_{\rm th}$,f), $3 \le Z \le 6$.}
\label{App:2}

We investigate the heavier isotopes $(Z > 2)$. Data for the binding energy, degeneracy, threshold energy, excited states and their decay modes are taken from the data tables \cite{nuclei}. Similar as for the H, He isotopes, we estimate the relevant, primary distribution by different approximations to treat $H$,
which leads to different values for the prefactor $R_{A,Z}(\lambda_T)$. The value $\lambda_T=1.3$ MeV was taken. In the approximation $H^{(0)}$ \cite{rnp20}, 
i.e. the sum over all bound states and neglecting interaction between them, we find unstable nuclei which decay to the final  states 
which are stable with respect to strong interaction.
We have excited states below the threshold $E^{\rm thresh}_{A,Z}$ of particle emission which perform an internal transition with $\gamma$ ray emission. 
We introduce the multiplier $R^\gamma_{AZ}(\lambda_T)$ to replace the ground state contribution by the intrinsic partition function including these excited states.
For instance, the threshold for emission of $n$ is given by $S_n$ as an edge of the continuum of scattering states. 
States above the threshold of decay to other clusters feed the corresponding final yields. This part of the intrinsic partition function is treated separately.

As already discussed for H, He, we have also to consider the scattering states. This leads to a change in the contributions obtained for bound states, 
and, in particular, for the unbound states. We use the  Beth-Uhlenbeck relation, see \cite{rnp20}
\begin{equation}
\label{Rviri}
R^{\rm vir}_{AZ,i}(\lambda_T)= [1-e^{-(E^{\rm thresh}_{A,Z}-E_{AZ,i}^{(0)} )/\lambda_T}]   \Theta(E^{\rm thresh}_{A,Z}-E_{AZ,i}^{(0)} )+\frac{1}{\pi \lambda_T}
e^{-(E^{\rm thresh}_{A,Z}-E_{AZ,i}^{(0)} )/\lambda_T} \int_0^\infty dE e^{-E/\lambda_T}\delta_{AZ,i}(E)) 
\end{equation}
as prefactor for the contribution of the different channels $\{A,Z,i\}$, degeneracy $g_{AZ,i}$.
 
 We are not able here to evaluate the integral over the scattering phase shifts for all channels of interest. We give only an estimate
 in analogy of the isotopes $^2$H, $^4$H, $^5$He which have been discussed in Ref. \cite{R20}. 
 Effective energy $E^{\rm eff}(T)$ are derived there from known phase shifts, also used to calculate the virial coefficients. We find  for $^4$H  the value 
 $R^{\rm vir}_{^4{\rm H}}(1.3)=0.0606 \times 1.473 = 0.0894$, and  for $^5$He  the value 
 $R^{\rm vir}_{^5{\rm He}}(1.3)=0.704$. Together with the known virial coefficient in the deuteron channel, 
in \cite{rnp20} the interpolation formula (in units of MeV)
\begin{equation}
\label{fitRvir}
R_{A,Z}^{\rm vir}(\lambda_T)=\frac{1}{e^{-(E^{\rm thresh}_{A,Z}+1.129)/0.204}+1}\,\,\frac{1}{e^{-(E^{\rm thresh}_{A,Z}+2.45)/\lambda_T}+1}
\end{equation}
has been introduced which reproduces the values for $^2$H, $^4$H, and $^5$He at $\lambda_T=1.3$ MeV. Values for H, He isotopes are given in the previous section.

In a subsequent step, the interaction between the constituents of nuclear matter has to be taken into account. 
In the low-density region considered here, we have self-energy shifts and Pauli blocking, a quasiparticle description can be introduced.
The generalized Beth-Uhlenbeck formula \cite{RMS82,SRS} uses quasiparticle energies. To avoid double counting, the scattering phase shift term is modified. Bound state energies are shifted by self-energy terms (also the continuum states so that it is nearly compensated in the binding energies) and Pauli blocking (which has to be included together with Hartree-Fock shifts in conserving approximations).
In the present work, we give only a discussion of these effects, see also \cite{R20}.\\

\subsubsection{Lithium}

\begin{table}
\begin{center}
 \begin{tabular}{|c|c|c|c|c|c|c|c|c|c|c|c|}
\hline
isotope& $A$  &  $Z$ & $Y^{\rm obs}_{A,Z}$&  $\frac{B_{A,Z}}{A} $&  $ g_{A,Z} $ & $Y^{\rm final}_{A,Z}$ & $E^{\rm thresh}_{AZ}$ & $E\,\, [g]$ & $R_{A,Z}^{\rm vir}(1.3)$ & $Y^{\rm rel, vir}_{A,Z}$& $\tilde Y^{\rm rel,vir}_{A,Z}$\\
\hline
 $\lambda_T$ & -&-&-&-&-		 & 0.968657&
 -&-&-& 0.950192&1.33495\\
$\lambda_n$ 	& -&-&-&-&-		 & -3.01812&
-&-&-& -3.01927&-2.97928 \\
$\lambda_p$ 	& -&-&-&-&-		 & -13.8252&
-&-&-&-14.1743&-17.9577 \\
\hline
6Li		&6 & 3 	& -		& 5.332	& 3 		&0.070551	&1.475  & -                     &0.953 	&0.0329	& 0.1046	  \\
7Li$^{\rm obs}$	&7 & 3 	& 6.7		& 5.606	& 4 		&9.358		& -	& -			&- 	&7.144 	&6.695	 \\
7Li		&7 & 3 	& -		& 5.606	& 4 		&{\it (9.358)}	& 2.461	& 0.478  [2]		&1.312  &6.5375 &5.931	 \\
8Li$^{\rm obs}$	&8 & 3 	& 4.2		& 5.160 & 5 		&5.196 		& -     & -			&- 	&3.3434 &4.208	\\
8Li		&8 & 3 	& -		& 5.160 & 5 		&{\it (5.196)}	& 2.038 & 0.980 [3]		&1.234  &3.3434 &4.208	\\
8Li*		&8 & 3 	& -		& 5.160 & (5) 		& -		& 2.038 & 2.255 [7], 3.210 [3] 	&0.224  &0.6069 &0.764	 \\
9Li$^{\rm obs}$	&9 & 3 	& 8.3		& 5.038 & 4 		&14.572		& -    	& -     		&- 	&8.396 	&8.296	\\
9Li		&9 & 3 	& -		& 5.038 & 4 		&{\it (14.572)}	& 4.062 & 2.691 [2] (4.301) [?]	&1.053  &8.16046&7.7152	 \\
10Li		&10 & 3 & -		& 4.531	&{\it  (3)} 	&{\it (0.549)}	& -0.032& -                     &0.861  &0.2362 &0.5807	 \\
11Li$^{\rm obs}$&11& 3	& 0.0045	& 4.155 & 4 		&0.0563		& -     & -    			&- 	&0.0246 &0.1445	 \\
11Li		&11& 3	& -		& 4.155 & 4 		&{\it (0.0563)}	& 0.396 & (1.266) [?]      	&0.899  &0.02397&0.1346	\\
12Li		&12& 3	& -		& 3.792 &{\it  (3)}   	&{\it (0.00173)}& -0.201& -      		&0.841&0.000646&0.0099357\\
\hline
 \end{tabular}
\caption{Data of  lithium nuclei [units: MeV, fm]. Mass number  $A$, charge number $Z$, observed yields  $Y^{\rm obs}_{A,Z}$ \cite{Koester}. Ground state binding energy $B_{A,Z}$ and degeneracy $g_{A,Z}$, excitation energy $E$ and degeneracy $g$ as well as continuum threshold energy $E^{\rm thresh}_{AZ}$ according \cite{nuclei}.  Excited state multiplier for $\lambda_T=1.3$ MeV. 
A simple NSE fit using the observed yields gives the distribution $Y^{\rm final}_{A,Z}$. In parentheses: yields of unstable isotopes
which  feed the stable isotopes. Excited state multiplier $R_{A,Z}^{\rm vir}(\lambda_T)$ for $\lambda_T=1.3$ MeV. Fit to all observed isotopes after feeding: $Y^{\rm rel, vir}_{A,Z}$. Fit excluding the weekly bound isotope $^{11}$Li gives $\tilde Y^{\rm rel, vir}_{A,Z}$.
 $^7$Li: $E^{\rm thresh}_{AZ}$  for decay to $^4$He+$^3$H; excited states at 4.630 MeV [8], 6.680 MeV [6] decay this way.  
}
\label{tab:Li}
\end{center}
\end{table}

The yields of Li isotopes are shown in Tab. \ref{tab:Li}. We consider $6 \le A \le 12$, in addition we denote with $^8$Li$^*$ the excited states of $^8$Li above $E^{\rm thresh}_{8,3}$ which decay to $^7$Li + $n$. The degeneracy is not known for $^{10}$Li and  $^{12}$Li, the value 3 was assumed, but is not of relevance for the discussion here.
The same holds also for some excited states.

We calculate the yields $Y^{\rm final}_{A,Z}$ relative to $Y_\alpha = 10000$ with only the observed ground states, 
optimizing the Lagrange parameters $\lambda_i$ to approximate the observed yields $Y^{\rm obs}_{A,Z}$ 
by the calculated yields of these ground states, i.e. taking the prefactor $R = 1$ in Eq. (3) of the letter \cite{rnp21}. 
As shown in Tab. \ref{tab:Li}, these final yields $Y^{\rm final}_{A,Z}$ agree not well with the observed ones, 
and we expect that also unstable nuclei should appear as given in italic parentheses.
In addition, there is a relative yield for $^6$Li which is small and below the limit of experimental sensitivity \cite{Koester,Sarah}, 
as well as yields for unstable nuclei $^{10}$Li and $^{12}$Li.
We conclude that the interpretation of the observed yields by a nuclear statistical equilibrium distribution remains unsatisfactory.

However, we have to construct within the information theoretical approach and the Lagrange parameters 
$\lambda_i$ not the final distribution, but the relevant (primary) distribution which evolves to the final 
distribution as described by reaction kinetics. In a next step we consider also the unstable and excited 
states which decay to the corresponding ground states, including continuum states. The threshold energy $E^{\rm thresh}_{AZ}$ is given in 
general by the neutron separation energy, 
but lower threshold energies appear for $^6$Li $\to \alpha + d$ and $^7$Li $\to \alpha + t$, see 
\cite{Jesinger05}. The corresponding 
prefactors $R^{\rm vir}_{AZ}(\lambda_T)$ are calculated for $\lambda_T=1.3$ MeV. The resulting yields $Y^{\rm rel,vir}_{A,Z}$ 
are given in Tab. \ref{tab:Li}, in addition are shown the final yields at the lines with the observed isotopes
which are the sum of the feeding contributions. 
$^7$Li is fed by $^8$Li*, $^{9}$Li by $^{10}$Li, and $^{11}$Li by $^{12}$Li. 
For other isotopes, the separation energy $S_n$ is large compared to $\lambda_T$  so that the feeding contributions are neglected.
The agreement with the observed yield becomes better. Only for $^{11}$Li a large discrepancy remains, outside the experimental errors.

The unsatisfactory coincidence with the observed yields is related to the overestimation of the observed yields for $^{11}$Li. 
If we drop this isotope, the determination of the Lagrange parameters is more in coincidence with the Lagrange parameters for H and He, 
they give the yields $\tilde Y^{\rm rel, vir}_{A,Z}$ which indicate the overestimation of the observed yields for $^{11}$Li. Note that within the virial approximation the Lagrange parameter $\lambda_p$ cannot be determined from the yields of only a fixed $Z$. However, because we consider the relation to the final yield of $\alpha$ particles, we can also determine $\lambda_p$. 
 
 As discussed in the case of $^6$He, weekly bound states are more sensitive to medium effects. 
The threshold energy of $^{11}$Li is low, and the yield is strongly overestimated within the different approaches. Obviously, the measured yield of $^{11}$Li is not well described by the relevant distribution in the approximation where the interaction between the components 
 is neglected.
Assuming that the medium effects are more important for the weakly bound states, the virial form of the intrinsic partition function is taken for the bound states  $^7$Li, $^8$Li, $^9$Li with large threshold energies, and an effective, relevant distribution $\tilde Y^{\rm rel, eff}_{A,Z}$ 
is calculated. Then, we can calculate the effective $R^{\rm eff}_{^{11}{\rm Li}}(1.3)=0.02799$. In-medium energy shifts may contribute to explain the effective reduction of $R$.

\subsubsection{Beryllium}

 \begin{table}
\begin{center}
 \begin{tabular}{|c|c|c|c|c|c|c|c|c|c|c|c|}
\hline
isotope& $A$  &  $Z$ & $Y^{\rm obs}_{A,Z}$&  $\frac{B_{A,Z}}{A} $&  $ g_{A,Z} $ & $Y^{\rm final}_{A,Z}$ & $E^{\rm thresh}_{AZ}$ & $E\,\,[g]$  & $R_{A,Z}^{\rm vir}(1.3)$ & $Y^{\rm rel, vir}_{A,Z}$ & $\tilde Y^{\rm rel, vir}_{A,Z}$ \\
\hline
 $\lambda_\beta$ & -&-&-&-&-		 & 0.970738&-&-&-& 0.92951&0.964348 \\
$\lambda_n$ 	& -&-&-&-&-		 & -3.34067&-&-&-& -3.18956&-3.13084 \\
$\lambda_p$ 	& -&-&-&-&-		 & -14.9358&-&-&-&-15.2619&-15.5065\\
\hline
7Be		&7 &4 	& $< 0.02$	& 5.372	& 4 	&{\it (1.88e-6)}& 1.585 & 0.429  [2]  			& 1.295	& 4.379e-7 	&7.145e-7\\
8Be		&8 &4 	& -		& 7.062	& 1 	&{\it (5.204)}	& -0.088& 3.03 [5]                      & 0.855 & 1.972		&1.857 \\
9Be$^{\rm obs}$	&9 &4 	& 4.4		& 6.462	& 4 	&4.406		& -  	& -    				& -	& 4.3179	&4.4\\
9Be		&9 &4 	& -		& 6.462	& 4 	&{\it (4.406)}	& 1.558 & -     			& 0.957	& 2.0369 	&2.16366\\
10Be$^{\rm obs}$&10&4 	& 46	 	& 6.497 & 1 	&46.102		& -  	& -           			& -	& 45.9604 	&46\\
10Be		&10&4 	& -	 	& 6.497 & 1 	&{\it (46.102)}	& 6.497 & 3.368 [5]   			& 1.369	& 42.0296 	&41.2161\\
10Be**		&10&4 	& -	 	& 6.497 & 1 	&-		& 6.497 & 5.958 [5], 5.959 [3]   	& 0.0743& 2.281 	&2.23693\\
11Be$^{\rm obs}$&11& 4	& 5.9		& 5.953 & 2 	&5.778		& -  	& -        			& -	& 6.34944   	&7.34292\\
11Be		&11& 4	& -		& 5.953 & 2 	&{\it (5.778)}	& 0.502 & 0.320 [2]         		& 1.596	& 6.34944   	&7.34292\\
11Be*		&11& 4	& -		& 5.953 & (2) 	&-		& 0.502 & 1.783 [6], 2.654 [4]        	& 0.1751& 0.6966   	&0.8056\\
12Be$^{\rm obs}$&12&4	& 2.8		& 5.721	& 1 	&2.758		& -  	& - 				& -	& 2.2599 	&2.8\\
12Be		&12&4	& -		& 5.721	& 1 	&{\it (2.758)}	& 3.170 & - 				& 0.9869& 2.18787 	&2.6912\\
12Be**		&12&4	& -		& 5.721	& (1) 	&-		& 3.170 &2.109 [5], 2.251 [1], 2.715 [3]& 1.4589& 3.23426 	&3.97831\\
13Be		&13&4	& -		& 5.241	& 2 	&{\it (0.1167)}	& -0.510& -				& 0.779	& 0.07206 	&0.108812\\
14Be$^{\rm obs}$&14&4	& 0.0027	& 4.994 & 1 	&0.0133		& -  	& -   				& -	& 0.010939 	&0.01853\\
14Be		&14&4	& -		& 4.994 & 1 	&{\it (0.0133)}	& 1.264 & -   				& 0.946	& 0.010931 	&0.0185174\\
15Be		&15&4	& -		& 4.541	& 6 	&{\it(0.000437)}&-1.800	& -   				& 0.0224& 7.90e-6 	&0.00001728\\
\hline
 \end{tabular}
\caption{Data of  beryllium nuclei [units: MeV, fm]. 
Mass number  $A$, charge number $Z$, observed yields  $Y^{\rm obs}_{A,Z}$ \cite{Koester}. Ground state binding energy $B_{A,Z}$ and degeneracy $g_{A,Z}$, excitation energy $E$ and degeneracy $g$ as well as continuum threshold energy $E^{\rm thresh}_{AZ}$ according \cite{nuclei}.  A simple NSE fit using the observed yields gives the distribution $Y^{\rm final}_{A,Z}$. In parentheses: yields of unstable isotopes
which  feed the stable isotopes. Excited state multiplier $R_{A,Z}^{\rm vir}(\lambda_T)$ for $\lambda_T=1.3$ MeV. Fit to all observed isotopes after feeding: $Y^{\rm rel, vir}_{A,Z}$. Fit excluding the weekly bound isotopes $^{11}$Be and $^{14}$Be gives $\tilde Y^{\rm rel, vir}_{A,Z}$.
}
\label{tab:Be}
\end{center}
\end{table}

Results for Be isotopes are shown in Tab. \ref{tab:Be}. For the relevant, primary distribution we have to consider also the unstable, excited states. Excited states above the continuum edge which emit a neutron are taken separately, their contribution to the intrinsic partition function is denoted by an asterisk. Also excited states little below the continuum edge may become dissolved in a dense medium if the binding energy is reduced owing to Pauli blocking, their contribution is denoted by two asterisks.
The final distribution is obtained with the feeding processes $^9$Be+$^{10}$Be** $\to ^9$Be$^{\rm final,vir}$,  
$^{10}$Be+$^{11}$Be*+$^{12}$Be** $\to ^{10}$Be$^{\rm final,vir}$, $^{12}$Be+$^{13}$Be $\to ^{12}$Be$^{\rm final,vir}$, 
$^{14}$Be+$^{15}$Be $\to ^{14}$Be$^{\rm final,vir}$. Because $^{11}$Be is only weakly bound we assume that $^{12}$Be** emits two neutrons.

Comparing the final distribution obtained from $Y^{\rm rel, vir}_{A,Z}$ with the observed distribution, we see once more that the weakly bound isotopes 
$^{11}$Be and $^{14}$Be are overestimated. As before, we assume that these weakly bound nuclei are strongly modified by medium effects and drop them 
during the fit procedure. The corresponding fit considering only the observed yields of $^{9}$Be, $^{10}$Be and $^{12}$Be 
is denoted by $\tilde Y^{\rm rel, vir}_{A,Z}$.

We can also introduce an effective intrinsic partition function and discuss the effects of medium modification and pairing. 
Note that the situation becomes more complex with respect to the faith of the excited states or the continuum states, and the discussion of the 
excited and continuum states gives some uncertainties. Fortunately, these highly excited states have only a very small effect.

\subsubsection{Boron}

 \begin{table}
\begin{center}
 \begin{tabular}{|c|c|c|c|c|c|c|c|c|c|c|c|}
\hline
isotope& $A$  &  $Z$ & $Y^{\rm obs}_{A,Z}$&  $\frac{B_{A,Z}}{A} $&  $ g_{A,Z} $ & $Y^{\rm final}_{A,Z}$ & $E^{\rm thresh}_{AZ}$ & $E\,\, [g]$  & $R_{A,Z}^{\rm vir}(1.3)$ & $Y^{\rm rel, vir}_{A,Z}$ & $\tilde Y^{\rm rel, vir}_{A,Z}$ \\
\hline
 $\lambda_\beta$ & -&-&-&-&-		 & 0.899772&-&-&-& 0.925253&0.874916 \\
$\lambda_n$ 	& -&-&-&-&-		 & -3.93373&-&-&-&-3.85808 &-3.78493 \\
$\lambda_p$ 	& -&-&-&-&-		 & -14.1993&-&-&-&-14.5279&-14.4323 \\
\hline
10B		&10 & 5 & $< 0.03$	& 6.475 & 7   &{\it (0.000605)}	& 4.466	& 0.718  [3], 1.740 [1], 2.154 [3], 3.587 [5]   & 1.398	& 0.0006285 	&0.000352 \\
11B$^{\rm obs}$	&11 & 5 & 1.6		& 6.928 & 4   &1.711		&- 	& -   						& -	& 1.618		& 1.6 \\
11B		&11 & 5 & -		& 6.928 & 4   &{\it (1.711)}	&8.674 	& 2.124  [2], 4.445 [6], 5.020 [4], 6.741 [8]   & 1.178	& 1.28901 	&1.25976 \\
12B$^{\rm obs}$	&12 & 5 & 1.0		& 6.631 & 3   &0.776		&-  	& -  						& -	&0.9708		& 1.0\\
12B		&12 & 5 & -		& 6.631 & 3   &{\it (0.776)}	&3.369  & 0.953 [5]  					& 1.771	&0.9708		& 1.0\\
12B*		&12 & 5 & -		& 6.631 & 3   &{\it (0.776)}	&3.369  &  1.674 [5], 2.621 [3], 2.723 [1]  		& 0.6019&0.32991 	& 0.339935\\
13B		&13 & 5 & -		& 6.496	& 4   &{\it (3.325)}	& 4.879 &3.482 [4?], 3.535 [4?], 3.681 [4?], 3.712 [4?] & 1.234	& 3.0558 	&3.646 \\
14B$^{\rm obs}$	&14 & 5 & 0.13		& 6.102 & 5   &0.1716		& - 	&- 						& -	& 0.1417 	&0.1536\\
14B		&14 & 5 & -		& 6.102 & 5   &{\it (0.1716)}	& 0.970 &- 						& 0.933	& 0.1417 	&0.1536\\
14B*		&14 & 5 & -		& 6.102 & (5) & -		& 0.970 &0.740 [3], 1.38 [7]				& 0.6902& 0.1048	&0.1136\\
15B$^{\rm obs}$	&15 & 5 & 0.046		& 5.880 & 4   &0.0425		& - 	& -						& -	& 0.04167 	& 0.0459\\
15B		&15 & 5 & -		& 5.880 & 4   &{\it (0.0425)}	& 2.78 	& (?)						& 0.982	& 0.04153 	& 0.04579\\
16B		&16 & 5 & -		& 5.507	& 1   &{\it (0.000134)}	&-0.082 & (?)						& 0.856	& 0.0001401 	& 0.000131\\
17B		&17 & 5 & $< 0.001$	& 5.270 & 4   &{\it (0.0000383)}& 1.39 	& (?)						& 0.95	& 0.000052 	&0.000052 \\
18B		&18 & 5 &-		& 4.977	& 5   &{\it (6.56e-7)}	&-0.005 & (?)						& 0.864	& 9.917e-7 	& 7.45e-7\\
\hline
 \end{tabular}
\caption{Data of  boron nuclei [units: MeV, fm]. 
Mass number  $A$, charge number $Z$, observed yields  $Y^{\rm obs}_{A,Z}$ \cite{Koester}. Ground state binding energy $B_{A,Z}$ and degeneracy $g_{A,Z}$, excitation energy $E$ and degeneracy $g$ as well as continuum threshold energy $E^{\rm thresh}_{AZ}$ according \cite{nuclei}.  A simple NSE fit using the observed yields gives the distribution $Y^{\rm final}_{A,Z}$. In parentheses: yields of unstable isotopes
which  feed the stable isotopes. Excited state multiplier $R_{A,Z}^{\rm vir}(\lambda_T)$ for $\lambda_T=1.3$ MeV. Fit to all observed isotopes after feeding: $Y^{\rm rel, vir}_{A,Z}$. Fit excluding the weekly bound isotope  $^{14}$B gives $\tilde Y^{\rm rel, vir}_{A,Z}$.}
\label{tab:B}
\end{center}
\end{table}

Results for B isotopes are shown in Tab. \ref{tab:B}, the coincidence of observed and calculated yields is rather good.
As before, we denoted the weakly bound part of the intrinsic partition function with $^{12}$B*, $^{14}$B*. 
It is not very essential whether the exited states are given to $^{11}$B, $^{13}$B as neutron decaying or not. 
In-medium effects may shift these states to the continuum.
$^{16}$B feeds $^{15}$B, and $^{18}$B feeds $^{17}$B.
If we drop the weakly bound nucleus $^{14}$B (we have already taken the excited states away), the fit to the other observed yields makes the overestimation more clear. Note that the yield of $^{13}$B is not obtained from the experiment. 
It is expected that continuum correlations and in-medium effects may give further corrections to describe the final yields of the isotopes.

\subsubsection{Carbon}

 \begin{table}
\begin{center}
 \begin{tabular}{|c|c|c|c|c|c|c|c|c|c|c|c|}
\hline
isotope& $A$  &  $Z$ & $Y^{\rm obs}_{A,Z}$&  $\frac{B_{A,Z}}{A} $&  $ g_{A,Z} $ & $Y^{\rm final}_{A,Z}$ & $E^{\rm thresh}_{AZ}$ & $E\,\, [g]$  & $R_{A,Z}^{\rm vir}(1.3)$ & $Y^{\rm rel, vir}_{A,Z}$ & $\tilde Y^{\rm rel, vir}_{A,Z}$ \\
\hline
 $\lambda_\beta$ & -&-&-&-&-		 & 0.924486&-&-&-& 0.723946&0.743479 \\
$\lambda_n$ 	& -&-&-&-&-		 & -3.50826&-&-&-&-3.49269 &-3.50412 \\
$\lambda_p$ 	& -&-&-&-&-		 & -15.939&-&-&-&-15.5599 &-15.587 \\
\hline
14C$^{\rm obs}$	&14 & 6 & 12.6	& 7.520 & 1 &13.923		& - 	& -					& -		& 13.0656	& 12.87  \\
14C		&14 & 6 & -	& 7.520 & 1 &{\it (13.923)}	& 8.176 & 6.093 [3], 6.589 [7]			& 1.0687	& 13.064	& 12.87  \\
15C$^{\rm obs}$	&15 & 6 & 4.3	& 7.100 & 2 &2.598		& - 	& -					& -		&4.003		&4.169  \\
15C		&15 & 6 & -	& 7.100 & 2 &{\it (2.598)}	& 1.218 & 0.740 [6]				& 2.47956	&2.91209 	&3.0686 \\
15C*		&15 & 6 & -	& 7.100 & 2 &-			& 1.218 &3.103 [2], 4.220 [6], 4.657 [4], 4.780 [4]& 0.001343	&0.00157	&0.00166 \\
16C$^{\rm obs}$	&16 & 6 & 5.0	& 6.922 & 1 &3.1999		& -	& -					& -		& 4.5398	&4.58291  \\
16C		&16 & 6 & -	& 6.922 & 1 &{\it (3.1999)}	& 4.250	& 1.766 [5]				& 2.4584	& 4.5394	&4.5824  \\
16C*		&16 & 6 & -	& 6.922 & 1 &-			& 4.250	& 3.986 [5], 4.089 [7], 4.142 [9]	& 0.59082	& 1.09095 	&1.10129 \\
17C$^{\rm obs}$	&17 & 6 & 0.64	& 6.558 & 4 &0.6973		& - 	& - 					& -		& 0.42058	&0.46212 \\
17C		&17 & 6 & -	& 6.558 & 4 &{\it (0.6973)}	& 0.734 & 0.217 [2], 0.332 [6] 			& 2.34904	& 0.42058	&0.46212 \\
17C*		&17 & 6 & -	& 6.558 & 4 &-			& 0.734 & 2.15 [8], 2.71 [2], 3.085 [10] 	& 0.002576	& 0.000461	&0.000506  \\
18C$^{\rm obs}$	&18 & 6 & 0.28	& 6.426 & 1 &0.3936		& -  	& -					& -		& 0.4059 	&0.4295  \\
18C		&18 & 6 & -	& 6.426 & 1 &{\it (0.3936)}	& 4.18  & 1.588 [5], 2.515 [5]			& 3.1311	& 0.3956 	&0.417553 \\
19C$^{\rm obs}$	&19 & 6 & 0.0025& 6.118 & 2 &0.0357		& - 	& -					& -		& 0.02233 	&0.0257997  \\
19C		&19 & 6 & -	& 6.118 & 2 &{\it (0.0357)}	& 0.580 & 0.209 [4], 0.282 [6]			& 4.5914	& 0.02233 	&0.0257997  \\
19C*		&19 & 6 & -	& 6.118 & 2 &-			& 0.580 & 0.653 [6], 1.46 [6]			& 2.1354	& 0.01038 	&0.01199  \\
20C$^{\rm obs}$	&20 & 6 & 0.0036& 5.961 & 1 &0.01087		& -  	& -   					& -		&  0.003034 	&0.003517 \\
20C		&20 & 6 & -	& 5.961 & 1 &{\it (0.01087)}	& 2.98  & 1.618 [5]    				& 2.3523	&  0.003034 	&0.003517  \\
\hline
 \end{tabular}
\caption{Data of  carbon nuclei [units: MeV, fm]. 
Mass number  $A$, charge number $Z$, observed yields  $Y^{\rm obs}_{A,Z}$ \cite{Koester}. Ground state binding energy $B_{A,Z}$ and degeneracy $g_{A,Z}$, excitation energy $E$ and degeneracy $g$ as well as continuum threshold energy $E^{\rm thresh}_{AZ}$ according \cite{nuclei}.  A simple NSE fit using the observed yields gives the distribution $Y^{\rm final}_{A,Z}$. In parentheses: yields of unstable isotopes
which  feed the stable isotopes. Excited state multiplier $R_{A,Z}^{\rm vir}(\lambda_T)$ for $\lambda_T=1.3$ MeV. Fit to all observed isotopes after feeding: $Y^{\rm rel, vir}_{A,Z}$. Fit excluding the weekly bound isotopes  $^{17}$C and $^{19}$C gives $\tilde Y^{\rm rel, vir}_{A,Z}$.}
\label{tab:C}
\end{center}
\end{table}

For carbon, the coincidence of inferred yields with the observed values is moderate, see Tab. \ref{tab:C}. As an anomaly, large deviation occur for $^{19}$C where the observed yield is overestimated by the calculated one.  $^{17}$C and $^{19}$C are weakly bound nuclei with the threshold of the continuum below 1 MeV. We expect that these weakly bound nuclei are stronger influenced by in-medium modifications, in particular Pauli blocking.
If we excluded both isotopes  from the least squared method, the resulting yields $\tilde Y^{\rm rel, vir}_{A,Z}$ show a significant suppression of the observed yields which may be discussed as a consequence of these medium modifications.
 
Calculations of the yields, see Figs. 3, 4 in \cite{Koestera}, cannot explain the low yield for $^{19}$C. Low yields of other 'exotic' nuclei ($^{11}$Li etc.) are also known, and are present  in ternary fission of other actinides (Am, Cf, etc.).
As claimed there, the authors have no explanation for the low yield of these exotic neutron rich, halo-like nuclei.

For the carbon isotopes, the spectrum of excited states is rather  complex and feeds different final states. In particular, isotopes with excited states above the 
neutron separation energy feed the yield of isotopes with $A-1$. The treatment of continuum states are also rather complex. 
To investigate the influence of medium effects, expressions for the Pauli blocking \cite{R20} may be used. However, the observed data are 
uncertain and the test of such effects is possible only after improved data are available.

\newpage

\section{Limits of the quasi-equilibrium primary distribution}

As explained in the letter, we calculate the yields $Y^{\rm rel, vir}_{A,Z}$ for a given set of Lagrange parameters $\lambda_i$, optimized to $A \le 10$,  and compare the final yields to the observed yields,
$Y^{\rm obs}_{A,Z}$/$Y^{\rm final, vir}_{A,Z}$,
shown in Fig. 1 of the Letter \cite{rnp21} for three actinides $^{241}$Pu($n_{\rm th}$,f), $^{253}$U($n_{\rm th}$,f),
 $^{245}$Cm($n_{\rm th}$,f).
%
%
The three Tables \ref{all241Pu}, \ref{all235U}, \ref{all245Cm} are given at the end. The values for $R_{A,Z}^{\rm vir}(1.3)$ are calculated according Eq. (\ref{fitRvir}).  A least squares fit of final yields $Y^{\rm final, vir}_{A,Z}$ to $Y^{\rm obs}_{A,Z}$ 
for $^2$H, $^3$H, $^4$He, $^8$He, $^7$Li,  $^8$Li,  $^9$Li has been performed to determine the Lagrange parameters $\lambda_i$.

 \begin{table}
\begin{center}
\hspace{0.5cm}
 \begin{tabular}{|c|c|c|c|c|c|c|c|c|c|c|}
\hline
isotope& $A$  &  $Z$ & $Y^{\rm obs}_{A,Z}$&  $\frac{B_{A,Z}}{A} $&  $ g_{A,Z} $  & $E^{\rm thresh}_{AZ}$ & $R_{A,Z}^{\rm vir}(1.3)$  & $Y^{\rm rel, vir}_{A,Z}$
& $Y^{\rm final, vir}_{A,Z}$ &$Y^{\rm obs}_{A,Z}$/$Y^{\rm final, vir}_{A,Z}$\\
\hline

0n	&1&0	&-		&0	& 2 	&-		&- 		& $1.5619  \times 10^6$	& $1.5619  \times 10^6$	& \\
1H	&1& 1 	& -		&0	& 2	&-              &-              & 19.1611	 	& 19.1611	 	& \\
2H 	&2& 1 	& 42 		& 1.112 & 3 	&2.224		& 0.973		& 41.431		& 41.431		& 1.01373\\
3H	&3& 1 	& 786		& 2.827 & 2 	&6.257		& 0.998		& 782.634 		& 786.927		& 0.99882 \\
4H	&4& 1 	& -		& 1.720	& 5 	&-1.6		& 0.0652	& 4.29277 		& [$\to ^3$H]		&	\\
\hline
3He	&3& 2 	&$< 0.01$	& 2.573	& 2 	& 5.494		& 0.997		& 0.00493465 		& 0.00493465 		&	\\
4He	&4& 2 	& 10000		& 7.073 & 1 	& 20.577	& 1 		& 8519.21 		& 10000 		& 1	\\
5He	&5& 2 	& -		& 5.512	& 4 	&-0.735		& 0.689 	& 1474.54 		& [$\to ^4$He]		& \\
6He	&6& 2 	& 260		& 4.878 & 1 	& 0.975		& 0.933 	& 225.52  		& 287.985 		& 0.902825	 \\
7He	&7& 2 	& -		& 4.123	& 4	& -0.410	& 0.875 	& 62.4646 		& [$\to ^6$He]		&	 \\
8He	&8& 2 	& 15		& 3.925 & 1 	& 2.125		& 0.971 	& 14.3563 		& 14.6189		& 1.02607\\
9He	&9& 2 	& -		& 3.349	& 2 	& -1.25		& 0.255 	& 0.262581		& [$\to ^8$He]		&	 \\
\hline
6Li	&6 & 3 	& -		& 5.332	& 3 	& 1.475     	& 0.953    	& 0.079027 		& 0.079027 		&	\\
7Li	&7 & 3 	& 6.7		& 5.606	& 4 	& 2.461		& 1.312		& 6.27821 		& 7.01482		&0.955121\\
8Li	&8 & 3 	& 4.2		& 5.160 & 5 	& 2.038     	& 1.234		& 4.05793	 	& 4.05793	 	& 1.03502\\
8Li*	&8 & 3 	& -		& 5.160 & (5) 	& 2.038     	& 0.224		& 0.736609		& [$\to ^7$Li]		&	\\
9Li	&9 & 3 	& 8.3		& 5.038 & 4 	& 4.062     	& 1.053 	& 8.00814  		& 8.47081 		&  0.979835	\\
10Li	&10& 3 	& -		& 4.531	&{\it 3}& -0.032    	& 0.861    	& 0.462669 		& [$\to ^9$Li]		&	\\
11Li	&11& 3	& 0.0045	& 4.155 & 4 	& 0.396     	& 0.899 	& 0.0852631  		& 0.090027 		& 0.049985	\\
12Li	&12& 3	& -		& 3.792 &{\it 3}& -0.201    	& 0.841 	& 0.0047639 		& [$\to ^{11}$Li]	&	\\
\hline
7Be	&7  & 4 &$< 0.02$	& 5.372	& 4 	& 1.585 	& 1.295		& 0.0000189218 		& 0.0000189218 		&	\\
8Be	&8  & 4 & -		& 7.062	& 1 	& -0.088 	& 1.270		& 3.12314	  	& [$\to ^4$He]		& \\
9Be	&9  & 4 & 4.4		& 6.462	& 4 	& 1.558  	& 0.957		& 3.69708 		& 3.69708 		& 1.19015	\\
10Be	&10 & 4 & 46	 	& 6.497 & 1 	& 6.497  	& 1.373		& 37.0544    		& 44.4526 		& 1.03481	\\
11Be	&11 & 4	& 5.9		& 5.953 & 2 	& 0.502  	& 1.596		& 12.5792 		& 12.5792 		& 0.469028	\\
11Be*	&11 & 4	& -		& 5.953 & (2) 	& 0.502  	& 0.00283	& 0.0223052		& [$\to ^{10}$Be]	&\\
12Be	&12 & 4	& 2.8		& 5.721	& 1 	& 3.170  	& 0.9869	& 5.10826 		& 5.59608		& 0.50035\\
12Be**	&12 & 4	& -		& 5.721	& (1) 	& 3.170  	& 1.425		& 7.37589 		& [$\to ^{10}$Be]	&	\\
13Be	&13 & 4	& -		& 5.241	& 2 	& -0.510 	& 0.779		& 0.487824 		& [$\to ^{12}$Be]	&	\\
14Be	&14 & 4	& 0.0027	& 4.994 & 1 	& 1.264  	& 0.946		& 0.121899   		& 0.122249 		& 0.022086	\\
15Be	&15 & 4	& -		& 4.541	& 6 	& -1.800 	& 0.0224	& 0.000350643 		& [$\to ^{14}$Be]	& 	\\
\hline
10B	&10 & 5 &  $< 0.03$	& 6.475 & 7   	& 4.466		& 1.381		& 0.00259095 		& 0.00259095 		&	\\
11B	&11 & 5 & 1.6		& 6.928 & 4  	& 8.674		& 1.178		& 1.65703	 	& 1.65703	 	& 0.965583\\
12B	&12 & 5 & 1.0		& 6.631 & 3  	& 3.369  	& 2.373		& 3.86715	 	& 3.86715	 	& 0.258588\\
13B	&13 & 5 & -		& 6.496	& 4   	& 4.879 	& 1.065		& 12.4409   		& 13.6133 		&	\\
14B	&14 & 5 & 0.13		& 6.102 & 5   	& 0.970 	& 1.234		& 3.71508 		& 3.71508 		& 0.0349925	\\
14B*	&14 & 5 & -		& 6.102 & (5)   & 0.970 	& 0.3894	& 1.17233 		& [$\to ^{13}$B]	&	\\
15B	&15 & 5 & 0.046		& 5.880 & 4   	& 2.78 		& 0.982		& 2.19877    		& 2.23925 		& 0.0205426	\\
16B	&16 & 5 &-		& 5.507	& 1   	& -0.082 	& 0.855		& 0.0404814 		& [$\to ^{15}$B]	&	\\
17B	&17 & 5 &$< 0.001$	& 5.270 & 4   	& 1.39 		& 0.95		& 0.055651   		& 0.0613263 		&	\\
18B	&18 & 5 &-		& 4.977	& 5   	& -0.005	& 0.864	 	& 0.0056753 		& [$\to ^{17}$B]	&	\\
\hline
14C	&14 & 6 & 12.6		& 7.520 & 1 	& 8.176		& 1.0687 	& 115.073  		& 115.146 		& 0.109426	\\
15C	&15 & 6 & 4.3		& 7.100 & 2 	& 1.218		& 2.47956	& 134.921   		& 185.177 		& 0.0232211	\\
15C*	&15 & 6 & -		& 7.100 & 2 	& 1.218		& 0.001343	& 0.0730938 		& [$\to ^{14}$C]	&	\\
16C	&16 & 6 & 5.0		& 6.922 & 1 	& 4.250		& 2.4584	& 209.121  		& 209.267 		& 0.0238929	\\
16C*	&16 & 6 & -		& 6.922 & 1 	& 4.250		& 0.59082	& 50.2557 		& [$\to ^{15}$C]	& 	\\
17C	&17 & 6 & 0.64		& 6.558 & 4 	& 0.734 	& 2.34904 	& 133.152 		& 133.152 		& 0.00480654	\\
17C*	&17 & 6 & -		& 6.558 & 4 	& 0.734 	& 0.002576 	& 0.146017 		& [$\to ^{16}$C]	&	\\
18C	&18 & 6 & 0.28		& 6.426 & 1 	& 4.18  	& 3.1311	& 129.411   		& 154.901 		& 0.00180761	\\
19C	&19 & 6 & 0.0025	& 6.118 & 2 	& 0.580 	& 4.5914	& 54.8071 		& 54.8071 		& 0.0000456145	\\
19C*	&19 & 6 & -		& 6.118 & 2 	& 0.580 	& 2.1354	& 25.49 		& [$\to ^{18}$C]	&		\\
20C	&20 & 6 & 0.0036	& 5.961 & 1 	& 2.98		& 2.3523	& 14.9109 		& 14.9109 		& 0.000241434	\\
\hline
 \end{tabular}
\caption{Observed yields of ternary fission of $^{241}$Pu($n_{\rm th}$,f) are compared to a final state distribution, obtained from feeding of a relevant (primary) distribution.
Least square fit of final yields $Y^{\rm final, vir}_{A,Z}$ to $Y^{\rm obs}_{A,Z}$ for $d,t,\alpha$, $^8$He, $^7$Li,  $^8$Li,  $^9$Li: Lagrange parameter values
$\lambda_T=1.2023,\,\,\lambda_n= -2.99811 ,\,\,\lambda_p=-16.6285$ MeV. }
\label{all241Pu}
\end{center}
\end{table}

 \begin{table}
\begin{center}
\hspace{0.5cm}
 \begin{tabular}{|c|c|c|c|c|c|c|c|c|c|c|}
\hline
isotope& $A$  &  $Z$ & $Y^{\rm obs}_{A,Z}$&  $\frac{B_{A,Z}}{A} $&  $ g_{A,Z} $  & $E^{\rm thresh}_{AZ}$ & $R_{A,Z}^{\rm vir}(1.3)$  & $Y^{\rm rel, vir}_{A,Z}$
& $Y^{\rm final, vir}_{A,Z}$ &$Y^{\rm obs}_{A,Z}$/$Y^{\rm final, vir}_{A,Z}$\\
\hline
0n	&1&0	&-		&0	& 2 	&-		&- 		& $1.3846  \times 10^6$	& $1.3846 \times 10^6$	& \\
1H	&1& 1 	& 115		&0	& 2	&-              &-              & 28.317	 	& 28.317	 	& \\
2H 	&2& 1 	& 50 		& 1.112 & 3 	&2.224		& 0.973		& 49.5345		& 49.5345		& 1.009\\
3H	&3& 1 	& 720		& 2.827 & 2 	&6.257		& 0.998		& 723.035 		& 728.398		& 0.9884 \\
4H	&4& 1 	& -		& 1.720	& 5 	&-1.6		& 0.0652	& 5.36206 		& [$\to ^3$H]		&	\\
\hline
3He	&3& 2 	&$< 0.01$	& 2.573	& 2 	& 5.494		& 0.997		& 0.00770 		& 0.00770 		&	\\
4He	&4& 2 	& 10000		& 7.073 & 1 	& 20.577	& 1 		& 8729.59 		& 10000 		& 1	\\
5He	&5& 2 	& -		& 5.512	& 4 	&-0.735		& 0.689 	& 1264.23 		& [$\to ^4$He]		& \\
6He	&6& 2 	& 191		& 4.878 & 1 	& 0.975		& 0.933 	& 157.339  		&  193.671		& 0.9862\\
7He	&7& 2 	& -		& 4.123	& 4	& -0.410	& 0.875 	& 36.3315 		& [$\to ^6$He]		&	 \\
8He	&8& 2 	& 8.2		& 3.925 & 1 	& 2.125		& 0.971 	& 6.73184 		& 5.58698		& 1.19962\\
9He	&9& 2 	& -		& 3.349	& 2 	& -1.25		& 0.255 	& 0.103633		& [$\to ^8$He]		&	 \\
\hline
6Li	&6 & 3 	& 0.05		& 5.332	& 3 	& 1.475     	& 0.953    	& 0.0895 		& 0.0895 		&	\\
7Li	&7 & 3 	& 4.1		& 5.606	& 4 	& 2.461		& 1.312		& 5.43439 		& 5.95131		&0.68896\\
8Li	&8 & 3 	& 1.8		& 5.160 & 5 	& 2.038     	& 1.234		& 2.8477	 	& 2.8477	 	& 0.63208\\
8Li*	&8 & 3 	& -		& 5.160 & (5) 	& 2.038     	& 0.224		& 0.516929		& [$\to ^7$Li]		&	\\
9Li	&9 & 3 	& 3.0		& 5.038 & 4 	& 4.062     	& 1.053 	& 4.45242  		& 4.66595 		& 0.642956\\
10Li	&10& 3 	& -		& 4.531	&{\it 3}& -0.032    	& 0.861    	& 0.213525 		& [$\to ^9$Li]		&	\\
11Li	&11& 3	& -		& 4.155 & 4 	& 0.396     	& 0.899 	& 0.0325  		& 0.0439449 		& 	\\
12Li	&12& 3	& -		& 3.792 &{\it 3}& -0.201    	& 0.841 	& 0.001510 		& [$\to ^{11}$Li]	&	\\
\hline
7Be	&7  & 4 &$< 0.01$	& 5.372	& 4 	& 1.585 	& 1.295		& 0.00002717 		& 0.00002717 		&	\\
8Be	&8  & 4 & 0.5		& 7.062	& 1 	& -0.088 	& 1.270		& 3.087	  		& [$\to ^4$He]		& 	\\
9Be	&9  & 4 & 2.9		& 6.462	& 4 	& 1.558  	& 0.957		& 2.97542 		& 2.97542 		& 1.19015\\
10Be	&10 & 4 & 32	 	& 6.497 & 1 	& 6.497  	& 1.373		& 24.0707    		& 27.1518 		& 1.178\\
11Be	&11 & 4	& 2.0		& 5.953 & 2 	& 0.502  	& 1.596		& 6.41261 		& 6.41261 		& 0.3118\\
11Be*	&11 & 4	& -		& 5.953 & (2) 	& 0.502  	& 0.00283	& 0.0223052		& [$\to ^{10}$Be]	&	\\
12Be	&12 & 4	& 1.5		& 5.721	& 1 	& 3.170  	& 0.9869	& 2.084 		& 2.25023		& 0.6665\\
12Be**	&12 & 4	& -		& 5.721	& (1) 	& 3.170  	& 1.425		& 3.08102 		& [$\to ^{10}$Be]	&	\\
13Be	&13 & 4	& -		& 5.241	& 2 	& -0.510 	& 0.779		& 0.166132 		& [$\to ^{12}$Be]	&	\\
14Be	&14 & 4	& -		& 4.994 & 1 	& 1.264  	& 0.946		& 0.033749   		& 0.0338313 		& - \\
15Be	&15 & 4	& -		& 4.541	& 6 	& -1.800 	& 0.0224	& 0.0000822353 		& [$\to ^{14}$Be]	& 	\\
\hline
10B	&10 & 5 &  $< 0.02$	& 6.475 & 7   	& 4.466		& 1.381		& 0.00259095 		& 0.00259095 		&	\\
11B	&11 & 5 & 0.25		& 6.928 & 4  	& 8.674		& 1.178		& 1.2522	 	& 1.2522	 	& 0.1996\\
12B	&12 & 5 & 0.17		& 6.631 & 3  	& 3.369  	& 2.373		& 2.33378	 	& 2.33378	 	& 0.07284\\
13B	&13 & 5 & 0.2		& 6.496	& 4   	& 4.879 	& 1.065		& 5.89343   		& 6.16591 		& 0.03243\\
14B	&14 & 5 & 0.1		& 6.102 & 5   	& 0.970 	& 1.234		& 1.44432 		& 1.44432 		& 0.06923\\
14B*	&14 & 5 & -		& 6.102 & (5)   & 0.970 	& 0.3894	& 0.272483 		& [$\to ^{13}$B]	&	\\
15B	&15 & 5 & -		& 5.880 & 4   	& 2.78 		& 0.982		& 0.68715    		& 0.69767 		& 0.0205426\\
16B	&16 & 5 &-		& 5.507	& 1   	& -0.082 	& 0.855		& 0.01052 		& [$\to ^{15}$B]	&	\\
17B	&17 & 5 &-		& 5.270 & 4   	& 1.39 		& 0.95		& 0.0117866   		& 0.01278 		&	\\
18B	&18 & 5 &-		& 4.977	& 5   	& -0.005	& 0.864	 	& 0.000997423 		& [$\to ^{17}$B]	&	\\
\hline
13C	&13 & 6 & 0.5		& 7.470 & 2 	& 4.946		& 1.0921 	& 1.95818  		& 1.95818 		& 0.25533\\
14C	&14 & 6 & 5.4		& 7.520 & 1 	& 8.176		& 1.0687 	& 59.763  		& 59.763 		& 0.090309\\
15C	&15 & 6 & 1.5		& 7.100 & 2 	& 1.218		& 2.47956	& 57.3539   		& 74.2395		& 0.0202049\\
15C*	&15 & 6 & -		& 7.100 & 2 	& 1.218		& 0.001343	& 0.0730938 		& [$\to ^{14}$C]	&	\\
16C	&16 & 6 & 0.2		& 6.922 & 1 	& 4.250		& 2.4584	& 70.2608  		& 70.3012 		& 0.002844\\
16C*	&16 & 6 & -		& 6.922 & 1 	& 4.250		& 0.59082	& 16.8856 		& [$\to ^{15}$C]	& 	\\
17C	&17 & 6 & -		& 6.558 & 4 	& 0.734 	& 2.34904 	& 36.8121 		& 36.8121 		& 	\\
17C*	&17 & 6 & -		& 6.558 & 4 	& 0.734 	& 0.002576 	& 0.0403688 		& [$\to ^{16}$C]	&	\\
18C	&18 & 6 & -		& 6.426 & 1 	& 4.18  	& 3.1311	& 28.3069   		& 32.9032 		& 	\\
19C	&19 & 6 & -		& 6.118 & 2 	& 0.580 	& 4.5914	& 9.88268 		& 9.88268 		& 	\\
19C*	&19 & 6 & -		& 6.118 & 2 	& 0.580 	& 2.1354	& 4.59631 		& [$\to ^{18}$C]	&	\\
20C	&20 & 6 & -		& 5.961 & 1 	& 2.98		& 2.3523	& 2.15661 		& 2.15661 		& 	\\
\hline
 \end{tabular}
\caption{Observed yields of ternary fission of $^{235}$U($n_{\rm th}$,f) are compared to a final state distribution, 
obtained from feeding of a relevant (primary) distribution.
Least square fit of final yields $Y^{\rm final, vir}_{A,Z}$ to $Y^{\rm obs}_{A,Z}$ for $d,t,\alpha$, $^8$He, $^7$Li,  $^8$Li,  $^9$Li: 
Lagrange parameter values
$\lambda_T=1.21899,\,\,\lambda_n= -3.2672 ,\,\,\lambda_p=-16.458$ MeV, see Tab. 1 of \cite{rnp21}. }
\label{all235U}
\end{center}
\end{table}

 \begin{table}
\begin{center}
\hspace{0.5cm}
 \begin{tabular}{|c|c|c|c|c|c|c|c|c|c|c|}
\hline
isotope& $A$  &  $Z$ & $Y^{\rm obs}_{A,Z}$&  $\frac{B_{A,Z}}{A} $&  $ g_{A,Z} $  & $E^{\rm thresh}_{AZ}$ & $R_{A,Z}^{\rm vir}(1.3)$  & $Y^{\rm rel, vir}_{A,Z}$
& $Y^{\rm final, vir}_{A,Z}$ &$Y^{\rm obs}_{A,Z}$/$Y^{\rm final, vir}_{A,Z}$\\
\hline
0n	&1&0	&-		&0	& 2 	&-		&- 		& 808957	& 808957& \\
1H	&1& 1 	& -		&0	& 2	&-              &-              & 17.5277	 	& 17.5277	 	& \\
2H 	&2& 1 	& - 		& 1.112 & 3 	&2.224		& 0.973		& 38.6885		& 38.6885		& \\
3H	&3& 1 	& 679		& 2.827 & 2 	&6.257		& 0.998		& 672.728 		& 679.378		& 0.9994 \\
4H	&4& 1 	& -		& 1.720	& 5 	&-1.6		& 0.0652	& 6.64915 		& [$\to ^3$H]		&	\\
\hline
3He	&3& 2 	&$< 0.6$	& 2.573	& 2 	& 5.494		& 0.997		& 0.007623 		& 0.007623 		&	\\
4He	&4& 2 	& 10000		& 7.073 & 1 	& 20.577	& 1 		& 8388.28 		& 10000 		& 1	\\
5He	&5& 2 	& -		& 5.512	& 4 	&-0.735		& 0.689 	& 1599.08 		& [$\to ^4$He]		& \\
6He	&6& 2 	& 286		& 4.878 & 1 	& 0.975		& 0.933 	& 252.97  		&  329.512		& 0.867951\\
7He	&7& 2 	& -		& 4.123	& 4	& -0.410	& 0.875 	& 76.542 		& [$\to ^6$He]		&	 \\
8He	&8& 2 	& 19		& 3.925 & 1 	& 2.125		& 0.971 	& 17.8185 		& 18.1823		& 1.04498\\
9He	&9& 2 	& -		& 3.349	& 2 	& -1.25		& 0.255 	& 0.363761		& [$\to ^8$He]		&	 \\
\hline
6Li	&6 & 3 	&$< 0.3$	& 5.332	& 3 	& 1.475     	& 0.953    	& 0.1456 		& 0.1456 		&	\\
7Li	&7 & 3 	& 13.6		& 5.606	& 4 	& 2.461		& 1.312		& 10.3859 		& 11.6359		&1.1602\\
8Li	&8 & 3 	& 5.6		& 5.160 & 5 	& 2.038     	& 1.234		& 6.88604	 	& 6.88604	 	& 0.813239\\
8Li*	&8 & 3 	& -		& 5.160 & (5) 	& 2.038     	& 0.224		& 1.24998		& [$\to ^7$Li]		&	\\
9Li	&9 & 3 	& 13.6		& 5.038 & 4 	& 4.062     	& 1.053 	& 13.234  		& 14.0611 		& 0.9672\\
10Li	&10& 3 	& -		& 4.531	&{\it 3}& -0.032    	& 0.861    	& 0.827114 		& [$\to ^9$Li]		&	\\
11Li	&11& 3	& -		& 4.155 & 4 	& 0.396     	& 0.899 	& 0.16309  		& 0.17299 		& 	\\
12Li	&12& 3	& -		& 3.792 &{\it 3}& -0.201    	& 0.841 	& 0.0099003		& [$\to ^{11}$Li]	&	\\
\hline
7Be	&7  & 4 &-		& 5.372	& 4 	& 1.585 	& 1.295		& 0.000055931 		& 0.000055931 		&	\\
8Be	&8  & 4 & -		& 7.062	& 1 	& -0.088 	& 1.270		& 6.3197	  	& [$\to ^4$He]		& 	\\
9Be	&9  & 4 & 9.1		& 6.462	& 4 	& 1.558  	& 0.957		& 7.74844 		& 7.74844 		& 1.1744\\
10Be	&10 & 4 & 66	 	& 6.497 & 1 	& 6.497  	& 1.373		& 74.0831    		& 85.786 		& 0.76935\\
11Be	&11 & 4	& 8.1		& 5.953 & 2 	& 0.502  	& 1.596		& 25.5214 		& 25.5214 		& 0.31738\\
11Be*	&11 & 4	& -		& 5.953 & (2) 	& 0.502  	& 0.00283	& 0.0452541		& [$\to ^{10}$Be]	&	\\
12Be	&12 & 4	& 5.5		& 5.721	& 1 	& 3.170  	& 0.9869	& 10.327 		& 11.4073		& 0.482149\\
12Be**	&12 & 4	& -		& 5.721	& (1) 	& 3.170  	& 1.425		& 3.08102 		& [$\to ^{10}$Be]	&	\\
13Be	&13 & 4	& -		& 5.241	& 2 	& -0.510 	& 0.779		& 1.08026 		& [$\to ^{12}$Be]	&	\\
14Be	&14 & 4	& -		& 4.994 & 1 	& 1.264  	& 0.946		& 0.278619   		& 0.279526 		&  \\
15Be	&15 & 4	& -		& 4.541	& 6 	& -1.800 	& 0.0224	& 0.0009075 		& [$\to ^{14}$Be]	& 	\\
\hline
10B	&10 & 5 &  $< 0.3$	& 6.475 & 7   	& 4.466		& 1.381		& 0.0088407 		& 0.0088407 		&	\\
11B	&11 & 5 & 2.4		& 6.928 & 4  	& 8.674		& 1.178		& 4.49871	 	& 4.49871	 	& 0.5334\\
12B	&12 & 5 & 2.3		& 6.631 & 3  	& 3.369  	& 2.373		& 10.4094	 	& 10.4094	 	& 0.22095\\
13B	&13 & 5 & 2.2		& 6.496	& 4   	& 4.879 	& 1.065		& 31.9379   		& 35.1112 		& 0.062658\\
14B	&14 & 5 & 0.21		& 6.102 & 5   	& 0.970 	& 1.234		& 10.0561 		& 10.0561 		& 0.02088\\
14B*	&14 & 5 & -		& 6.102 & (5)   & 0.970 	& 0.3894	& 3.17331 		& [$\to ^{13}$B]	&	\\
15B	&15 & 5 & -		& 5.880 & 4   	& 2.78 		& 0.982		& 5.98911    		& 6.10871 		& 	\\
16B	&16 & 5 &-		& 5.507	& 1   	& -0.082 	& 0.855		& 0.119593 		& [$\to ^{15}$B]	&	\\
17B	&17 & 5 &-		& 5.270 & 4   	& 1.39 		& 0.95		& 0.17089   		& 0.189728 		&	\\
18B	&18 & 5 &-		& 4.977	& 5   	& -0.005	& 0.864	 	& 0.018838 		& [$\to ^{17}$B]	&	\\
\hline
13C	&13 & 6 & 1.1		& 7.470 & 2 	& 4.946		& 1.0921 	& 9.3179  		& 9.3179		& 0.11805\\
14C	&14 & 6 & 23.3		& 7.520 & 1 	& 8.176		& 1.0687 	& 329.638  		& 329.858 		& 0.070636\\
15C	&15 & 6 & 7.7		& 7.100 & 2 	& 1.218		& 2.47956	& 404.969   		& 551.125		& 0.013971\\
15C*	&15 & 6 & -		& 7.100 & 2 	& 1.218		& 0.001343	& 0.0730938 		& [$\to ^{14}$C]	&	\\
16C	&16 & 6 & 9.5		& 6.922 & 1 	& 4.250		& 2.4584	& 608.157  		& 608.607 		& 0.0156094\\
16C*	&16 & 6 & -		& 6.922 & 1 	& 4.250		& 0.59082	& 146.157 		& [$\to ^{15}$C]	& 	\\
17C	&17 & 6 & -		& 6.558 & 4 	& 0.734 	& 2.34904 	& 410.736 		& 410.736 		& 	\\
17C*	&17 & 6 & -		& 6.558 & 4 	& 0.734 	& 0.002576 	& 0.45042 		& [$\to ^{16}$C]	&	\\
18C	&18 & 6 & 0.32		& 6.426 & 1 	& 4.18  	& 3.1311	& 387.563   		& 468.869 		&0.00068249 	\\
19C	&19 & 6 & -		& 6.118 & 2 	& 0.580 	& 4.5914	& 174.818 		& 174.818 		& 	\\
19C*	&19 & 6 & -		& 6.118 & 2 	& 0.580 	& 2.1354	& 81.3057 		& [$\to ^{18}$C]	&	\\
20C	&20 & 6 & -		& 5.961 & 1 	& 2.98		& 2.3523	& 47.6247 		& 47.6247 		& 	\\
\hline
 \end{tabular}
\caption{Observed yields of ternary fission of $^{245}$Cm($n_{\rm th}$,f) are compared to a final state distribution, obtained from feeding of a relevant (primary) distribution.
Least square fit of final yields $Y^{\rm final, vir}_{A,Z}$ to $Y^{\rm obs}_{A,Z}$ for $t,\alpha$, $^8$He, $^7$Li,  $^8$Li,  $^9$Li: Lagrange parameter values
$\lambda_T=1.24059,\,\,\lambda_n= -2.9971 ,\,\,\lambda_p=-16.3586$ MeV. }
\label{all245Cm}
\end{center}
\end{table}